\documentclass[aps,prd,twocolumn,superscriptaddress,showpacs]{revtex4}
\usepackage[dvipdfmx]{graphicx}
\usepackage{here}
\usepackage{amsmath,amssymb,times}
\usepackage{color}
\usepackage{mathrsfs}

\newcommand{\bequ}{\begin{equation}}
\newcommand{\eequ}{\end{equation}}
\newcommand{\bea}{\begin{eqnarray}}
\newcommand{\eea}{\end{eqnarray}}

\renewcommand{\a}{\alpha}

\def\gsim{~\,\makebox(1,1){$\stackrel{>}{\widetilde{}}$}\,~}
\def\lsim{~\,\makebox(1,1){$\stackrel{<}{\widetilde{}}$}\,~}

\DeclareSymbolFont{boldletters}{OML}{cmm} {b}{it}
\DeclareSymbolFontAlphabet{\mathbit}{boldletters}
\DeclareMathSymbol{\alpha}{\mathalpha}{letters}{"0B}
\DeclareMathSymbol{\beta}{\mathalpha}{letters}{"0C}
\DeclareMathSymbol{\gamma}{\mathalpha}{letters}{"0D}
\DeclareMathSymbol{\delta}{\mathalpha}{letters}{"0E}
\DeclareMathSymbol{\epsilon}{\mathalpha}{letters}{"0F}
\DeclareMathSymbol{\zeta}{\mathalpha}{letters}{"10}
\DeclareMathSymbol{\eta}{\mathalpha}{letters}{"11}
\DeclareMathSymbol{\theta}{\mathalpha}{letters}{"12}
\DeclareMathSymbol{\iota}{\mathalpha}{letters}{"13}
\DeclareMathSymbol{\kappa}{\mathalpha}{letters}{"14}
\DeclareMathSymbol{\lambda}{\mathalpha}{letters}{"15}
\DeclareMathSymbol{\mu}{\mathalpha}{letters}{"16}
\DeclareMathSymbol{\nu}{\mathalpha}{letters}{"17}
\DeclareMathSymbol{\xi}{\mathalpha}{letters}{"18}
\DeclareMathSymbol{\pi}{\mathalpha}{letters}{"19}
\DeclareMathSymbol{\rho}{\mathalpha}{letters}{"1A}
\DeclareMathSymbol{\sigma}{\mathalpha}{letters}{"1B}
\DeclareMathSymbol{\tau}{\mathalpha}{letters}{"1C}
\DeclareMathSymbol{\upsilon}{\mathalpha}{letters}{"1D}
\DeclareMathSymbol{\phi}{\mathalpha}{letters}{"1E}
\DeclareMathSymbol{\chi}{\mathalpha}{letters}{"1F}
\DeclareMathSymbol{\psi}{\mathalpha}{letters}{"20}
\DeclareMathSymbol{\omega}{\mathalpha}{letters}{"21}
\DeclareMathSymbol{\varepsilon}{\mathalpha}{letters}{"22}
\DeclareMathSymbol{\vartheta}{\mathalpha}{letters}{"23}
\DeclareMathSymbol{\varpi}{\mathalpha}{letters}{"24}
\DeclareMathSymbol{\varrho}{\mathalpha}{letters}{"25}
\DeclareMathSymbol{\varsigma}{\mathalpha}{letters}{"26}
\DeclareMathSymbol{\varphi}{\mathalpha}{letters}{"27}
\DeclareMathSymbol{\Gamma}{\mathalpha}{letters}{"00}
\DeclareMathSymbol{\Delta}{\mathalpha}{letters}{"01}
\DeclareMathSymbol{\Theta}{\mathalpha}{letters}{"02}
\DeclareMathSymbol{\Lambda}{\mathalpha}{letters}{"03}
\DeclareMathSymbol{\Xi}{\mathalpha}{letters}{"04}
\DeclareMathSymbol{\Pi}{\mathalpha}{letters}{"05}
\DeclareMathSymbol{\Sigma}{\mathalpha}{letters}{"06}
\DeclareMathSymbol{\Upsilon}{\mathalpha}{letters}{"07}
\DeclareMathSymbol{\Phi}{\mathalpha}{letters}{"08}
\DeclareMathSymbol{\Psi}{\mathalpha}{letters}{"09}
\DeclareMathSymbol{\Omega}{\mathalpha}{letters}{"0A}

\begin{document}
\preprint{SAGA-HE-288}

\title{Crossover-model approach to QCD phase diagram, equation of state \\
and susceptibilities 
in the 2+1 and 2+1+1 flavor systems}

\author{Akihisa Miyahara}
\email[]{miyahara@email.phys.kyushu-u.ac.jp}
\affiliation{Department of Physics, Graduate School of Sciences, Kyushu University,
             Fukuoka 819-0395, Japan}

\author{Masahiro Ishii}
\email[]{ishii@phys.kyushu-u.ac.jp}
\affiliation{Department of Physics, Graduate School of Sciences, Kyushu University,
             Fukuoka 819-0395, Japan}

\author{Hiroaki Kouno}
\email[]{kounoh@cc.saga-u.ac.jp}
\affiliation{Department of Physics, Saga University,
             Saga 840-8502, Japan}

\author{Masanobu Yahiro}
\email[]{yahiro@phys.kyushu-u.ac.jp}
\affiliation{Department of Physics, Graduate School of Sciences, Kyushu University,
             Fukuoka 819-0395, Japan}             

\date{\today}

\begin{abstract}
We construct a simple model 
for describing the hadron-quark crossover transition 
by using lattice QCD (LQCD) data in the 2+1 flavor system, and draw the phase diagram 
in the 2+1 and 2+1+1 flavor systems 
through analyses of the equation of state (EoS) and the susceptibilities. 
In the present hadron-quark crossover (HQC) model, 
the entropy density $s$ is defined by 
$s=f_{\rm H}s_{\rm H}+(1-f_{\rm H})s_{\rm Q}$ with 
the hadron-production probability $f_{\rm H}$, 
where $s_{\rm H}$ is calculated by the hadron resonance gas model  
valid in low temperature ($T$) and $s_{\rm Q}$ is evaluated by 
the independent quark model that explains 
LQCD data on the EoS in the region 
$400 \lsim T \leq  500$~MeV for the 2+1 flavor systems and 
$400 \lsim T \leq  1000$~MeV for the 2+1+1 flavor system. 
The $f_{\rm H}$ is determined from 
LQCD data on $s$ and susceptibilities for the baryon-number ($B$), the isospin ($I$) and 
the hypercharge ($Y$) in the 2+1flavor system. 
The HQC model is successful in reproducing LQCD data on the EoS and 
the flavor susceptibilities $\chi_{ff'}^{(2)}$ 
for $f,f=$u, d, s in the 2+1+1 flavor system, 
without changing the $f_{\rm H}$. 
We define the hadron-quark transition temperature with $f_{\rm H}=1/2$. 
For the 2+1 flavor system, the transition line thus obtained 
is almost identical in $\mu_{B}$--$T$, $\mu_{I}$--$T$, $\mu_{Y}$--$T$ planes, when 
the chemical potentials $\mu_{\a}$ ($\a=B, I, Y$) are smaller than 250~MeV. 
This $BIY$ approximate equivalence persists also in the 2+1+1 flavor system. 
We plot the phase diagram also in $\mu_{\rm u}$--$T$, $\mu_{\rm d}$--$T$, 
$\mu_{\rm s}$--$T$, $\mu_{\rm c}$--$T$ planes in order to investigate 
flavor dependence of transition lines. 
In the 2+1+1 flavor system, c quark does not affect the 2+1 flavor 
subsystem composed of u, d, s. 
The flavor off-diagonal susceptibilities are good indicators 
to see how hadrons survive as $T$ increases, 
since the independent quark model hardly contributes to them. $T$ dependence of the off-diagonal susceptibilities and 
the $f_{\rm H}$ show that 
the transition region at $\mu_{\a}=0$ is $170 \lsim T \lsim 400$~MeV  
for both the 2+1 and 2+1+1 flavor systems. 
\end{abstract}

\maketitle

\section{Introduction}
\label{Introduction}
\vspace{-10pt}
Lattice QCD (LQCD) is the first-principle calculation of QCD, and 
has been providing a lot of information on hot QCD. 
Recently, the continuum and thermodynamic limits were taken in 
2+1 flavor LQCD simulations~\cite{YAoki_crossover}, and 
it was confirmed  that the chiral and deconfinement transitions 
are crossover at zero chemical potential.

As an  approach alternative to LQCD simulations, 
we can consider effective models. 
This approach is useful for the physical interpretation of LQCD data and 
the prediction of physical quantities that are difficult to calculate in LQCD simulations. 
Recently, the Polyakov-loop extended Nambu-Jona-Lasinio (PNJL) type models 
have been used extensively, since they can treat both the chiral and the 
deconfinement transitions~\cite{MO,Dumitru,Fukushima,Ratti,Megias,Rossner,Schaefer,Abuki,Kashiwa1,Sakai_EPNJL,Gatto_EPNJL,Sasaki_EPNJL,Kashiwa_nonlocal}. 
In the PNJL-type models, the pseudocritical temperature of 
the deconfinement transition is almost equal to or lower than that 
of the chiral transition. 
For the 2 flavor system, the PNJL-type models well explain 
LQCD data~\cite{Sakai_EPNJL,Gatto_EPNJL,Sasaki_EPNJL,Kashiwa_nonlocal}, 
since the two transitions take place almost simultaneously in LQCD 
simulations. 
For the 2+1 flavor system, however, LQCD shows that 
the chiral-transition temperature is considerably lower than the 
deconfinement-transition one~\cite{YAoki_crossover,YAoki_Tc}. 
For this reason, it is not easy for the PNJL-type models 
to explain the chiral and deconfinement transitions simultaneously. 
It is thus important to construct a reasonable effective model 
for the 2+1 flavor system.

Now our discussion moves to the 2+1 flavor system composed of 
up (u), down (d), strange (s) quarks. 
For low temperature ($T$), the hadron resonance gas (HRG) model 
reproduces LQCD data on the equation of state (EoS) and 
the baryon-number ($B$) 
susceptibility~\cite{Borsanyi_order,Borsanyi_sus,Steinheimer}.  
In addition, below the chiral-transition temperature,  
the absolute value of chiral condensate is explained 
by HRG+chiral perturbation theory ($\chi$PT)~\cite{Borsanyi_Tc}. 
Recently, it was reported that the HRG model 
also accounts for LQCD data on $T$ dependence of 
the Polyakov loop~\cite{Megias_PRL}.  
These results suggest that the hadron degree of freedom  
is important in QCD phase transitions, 
although it is not treated explicitly in the PNJL-type models. 

Recently, the EoS, the baryon-number susceptibilities $\chi_{B}^{(n)}$ 
($n=2\sim4$) and the isospin ($I$) susceptibility $\chi_{I}^{(2)}$  
were well described 
by hadron-quark hybrid models~\cite{Albright:2014gva,Miyahara:2016din}. 
This model also reproduces qualitatively that the chiral-transition 
temperature is lower than the deconfinement-transition 
one~\cite{Miyahara:2016din}. 
In the hadron-quark hybrid model of Ref.~\cite{Albright:2014gva}, 
the pressure $P$ is defined by 
\begin{eqnarray}
P&=&f(T,\{\mu_\a \}) P_{\rm H}(T,\{\mu_\a \}) 
\nonumber \\
&~& + \left[1-f(T,\{\mu_\a \}) \right] 
P_{\rm Q}(T,\{\mu_\a \}), 
\label{P_Hybrid_1}
\end{eqnarray}
where 
the $\mu_{\a}$ are the chemical potentials of quantum numbers 
$\a=B, I$ and hypercharge $Y$. 
The hadron piece $P_{\rm H}$ is  calculated by the HRG model with 
the excluded volume and the quark-gluon piece $P_{\rm Q}$ is  
evaluated with perturbative QCD~\cite{Albright:2014gva}. 
Here, the fraction factor $f(T,\{\mu_\a \})$ 
is determined so as 
to reproduce LQCD data on $P$ and the interaction measure. 
The other thermal quantities are obtainable from $P$.

Alternatively, one can start with the entropy density    
\begin{eqnarray}
s&=&f_{\rm H}(T,\{\mu_\a \}) s_{\rm H}(T,\{\mu_\a \}) 
\nonumber \\
&~& + \left[1-f_{\rm H}(T,\{\mu_\a \}) \right] 
s_{\rm Q}(T,\{\mu_\a \})  
\label{s_Hybrid_def}
\end{eqnarray}
and calculate the other thermal quantities 
from $s$~\cite{Hatsuda_hybrid model}. 
As an advantage of this approach, $T$ dependence of $s$ 
cannot be free. In fact, the $T$ dependence should satisfy 
the thermodynamic inequality and the Nernst's theorem~\cite{Landau-Lifshitz}: 
\bea
\left. \frac{\partial s(T,\{\mu_\a \})}{\partial T}\right|_{ {\{\mu_\a \}}=0 }> 0,
~~~
\left. s(T,\{\mu_\a \})\right|_{T={\{\mu_\a \}}=0} =0.  
\eea 
In this paper, this condition is automatically satisfied, since we use 
LQCD data as $s$.  
In Eq.~\eqref{s_Hybrid_def}, the factor $f_{\rm H}$ means 
the hadron-production probability. 
When $f_{\rm H}(T,{\{\mu_\a \}})=1~(0)$, the system is in the hadron (quark) 
phase composed of hadron (quark-gluon) matter only. 
In principle, the factor $f_{\rm H}$ is determinable from LQCD data on $s$, 
if $s_{\rm H}$ and $s_{\rm Q}$ are given. 
The hadron-quark hybrid model of Ref.~\cite{Miyahara:2016din} takes Eq.~\eqref{s_Hybrid_def}, and the hadron piece $s_{\rm H}$ is obtained 
by the HRG model and the quark-gluon piece $s_{\rm Q}$ is 
by the simple quark model in which an adjustable parameter is introduced  
so as to reproduce LQCD data \cite{Borsanyi:2016ksw} on $s(T,0)$ 
at $T =300$~MeV. 
The quark model is a simplified version of 
PNJL model: Namely, this model takes account of the coupling between the 
quark field and the homogeneous classical gauge field, but does not treat 
the quark-quark couplings that are expected to be 
not important above the chiral- and 
deconfinement-transition temperatures.  
We refer to the simple model as ''independent quark (IQ) model" 
and the hadron-quark hybrid model of Ref.~\cite{Miyahara:2016din} 
as ''hadron-quark crossover (HQC) model" in this paper.

Lately, state-of-art LQCD data on the EoS and 
the flavor diagonal and off-diagonal 
susceptibilities, $\chi_{ff'}$, became available 
for the 2+1+1 flavor system~\cite{Bellwied:2015lba, Borsanyi:2016ksw} 
in addition to the case of 
the 2+1 flavor system~\cite{Borsanyi_sus_plot, Borsanyi:2016ksw}, 
where $f$=u, d, s for the 2+1 system and $f$=u, d, s, charm (c) 
for the 2+1+1 flavor system. 
It is an interesting question how c quark 
behaves in the 2+1+1 flavor system.

When the current quark mass $m$ is infinity, 
$Z_3$ symmetry is exact and the Polyakov loop 
is an order parameter of the spontaneous $Z_3$ symmetry breaking.  
Dynamical quark with finite $m$ breaks $Z_3$ symmetry explicitly through the temporal 
boundary condition for quark.
For the 2+1 and 2+1+1 flavor systems, it is not clear that 
the Polyakov loop is still a good order parameter 
of the confinement-deconfinement (hadron-quark) transition. 
As a reasonable assumption, we define the hadron-quark  transition 
temperature by the condition $f_{\rm H}=1/2$. Another interesting question is 
how the phase diagram is 
in $\mu_{B}$--$T$, $\mu_{I}$--$T$, $\mu_{Y}$--$T$ planes and also 
in $\mu_{\rm u}$--$T$, $\mu_{\rm d}$--$T$, $\mu_{\rm s}$--$T$, $\mu_{\rm c}$--$T$ planes, 
where $\mu_{f}$ ($f=$u, d, s, c) is the chemical potential for $f$ quark.

In this paper, we reconstruct the HQC model 
by using new LQCD data~\cite{Borsanyi:2016ksw,Borsanyi_sus_plot} 
in the 2+1 flavor system, and draw the phase diagram 
in the 2+1 and 2+1+1 flavor systems 
through analyses of the equation of state (EoS) and the susceptibilities.

In the previous work of Ref.~\cite{Miyahara:2016din}, 
the IQ model had 
the momentum cutoff $\Lambda_{\rm T}$ in the thermal quark-loop term 
of ${P_Q}$. 
The cutoff $\Lambda_{\rm T}$ was introduced as an adjustable parameter 
to reproduce LQCD data \cite{Borsanyi_entropy} on $s$ at $T=300$~MeV, 
where the data were available in $T \leq 400$~MeV. 
Recently, however, we found that 
the IQ  model begins to underestimate new LQCD data 
\cite{Borsanyi:2016ksw} as $T$ increases from 400~MeV; 
here, LQCD data on $s$ were deduced from new LQCD data 
\cite{Borsanyi:2016ksw} on $P$ by differentiating $P$ 
with respect to $T$ and thereby LQCD data on $s$ became available in 
$T \leq 500$~MeV. 
We then reformulate the IQ model slightly 
so that the model can explain the new data 
in $400 \lsim T \leq  500$~MeV where 
the data is consistent with NNLO hard thermal loop 
(HTL) perturbation~\cite{Haque:2013sja}.

In the HQC model, $f_{\rm H}(T,{\{\mu_\a \}})$ is determined from 
LQCD data on $s$, 
the baryon-number susceptibility $\chi_{B}^{(2)}$, 
the isospin susceptibility $\chi_{I}^{(2)}$ and 
the hypercharge ($Y$) susceptibility $\chi_{Y}^{(2)}$ in the 
2+1 flavor system, where $\a=B, I, Y$. 
The HQC model with the $f_{\rm H}(T,{\{\mu_\a \}})$ 
automatically reproduces LQCD data on the EoS and the $\chi_{ff'}^{(2)}$ 
in the 2+1 flavor system. 
In particular, the off-diagonal susceptibilities $\chi_{ff'}^{(2)}$ 
($f \neq f'$) are good indicators to see 
how hadrons survive as $T$ increases, {since the IQ  model hardly contributes to the off-diagonal susceptibilities~\cite{Kouno_3f_FTBC}.}  
In practice, the upper limit of the transition region is clearly determined 
by the off-diagonal susceptibilities. 
We then determine the transition region for the 2+1 flavor system with 
zero chemical potential 
from $T$ dependence of $f_{\rm H}(T,0)$ and the off-diagonal susceptibilities.

Next, we draw the phase diagram 
in $\mu_{B}$--$T$, $\mu_{I}$--$T$, $\mu_{Y}$--$T$ planes. 
The transition lines are almost identical in these planes, 
when $\mu_{\a} < 250$~MeV. 
This property is referred to as ''$BIY$ approximate equivalence" in 
this paper. What is the nature of $BIY$ approximate equivalence? 
This is discussed. We also plot the phase diagram 
in $\mu_{\rm u}$--$T$, $\mu_{\rm d}$--$T$, $\mu_{\rm s}$--$T$ planes 
to see flavor dependence of hadron-quark transition lines.

The HQC model is applied to 
the EoS and the $\chi_{ff'}$ in the 2+1+1 flavor system 
without changing the $f_{\rm H}(T,{\{\mu_\a \}})$. 
The HQC model succeeds in reproducing LQCD data on the EoS and the 
$\chi_{ff'}^{(2)}$ for $f, f'=$u, d, s, and explains 
$\chi_{\rm cc}^{(2)}$ qualitatively. 
We then determine the transition region for the 2+1+1 flavor system with 
zero chemical potential 
from $T$ dependence of $f_{\rm H}(T,0)$ and the off-diagonal susceptibilities, 
and investigate the role of c quark in the 2+1+1 flavor system.

Finally, we draw the phase diagram 
in $\mu_{B}$--$T$, $\mu_{I}$--$T$, $\mu_{Y}$--$T$, $\mu_{Y}$--$T$ planes 
to see whether $BIY$ approximate equivalence persists in the 2+1+1 flavor 
system, and plot the diagram 
in $\mu_{\rm u}$--$T$, $\mu_{\rm d}$--$T$, $\mu_{\rm s}$--$T$, $\mu_{\rm c}$--$T$ planes 
to investigate flavor dependence of hadron-quark transition lines.

This paper is organized as follows. 
In Sec.~\ref{model}, we recapitulate the HRG model and 
reformulate the IQ model 
without the cutoff $\Lambda_{\rm T}$. 
We review the HQC model. 
Numerical results are shown in Sec~\ref{results}. 
Section~\ref{summary} is devoted to summary.

\section{Model building}
\label{model}
\vspace{-10pt}
We reformulate the hadron-quark crossover (HQC) model 
of Ref.~\cite{Miyahara:2016din}. 
This model consists of 
the hadron resonance gas (HRG) 
model reliable for small $T$ and the independent quark (IQ) model 
reasonable for large $T$. 

For later convenience, we define several kinds of chemical potentials. For the 2+1 flavor system, we represent the chemical potentials of u, d, s quarks by $\mu_{\rm u}, \mu_{\rm d}$ and $\mu_{\rm s}$, respectively. These potentials are related to the baryon-number ($B$) chemical potential $\mu_{B}$, the isospin ($I$) chemical potential $\mu_{I}$ and the hypercharge ($Y$) chemical potential $\mu_{Y}$ as
\begin{eqnarray}
\begin{array}{l}
\mu_{B} = \mu_{\rm u} + \mu_{\rm d} + \mu_{\rm s},\\
\mu_{I} = \mu_{\rm u} - \mu_{\rm d},\\
\mu_{Y} = \frac{1}{2}(\mu_{\rm u} + \mu_{\rm d} - 2\mu_{\rm s})
\end{array}
\label{chemical potential 2+1}
\end{eqnarray}
for the 2+1 flavor system. 
As for $\mu_{I}$ and $\mu_{Y}$, 
the right-hand side of Eq.~\eqref{chemical potential 2+1} 
stems from the diagonal elements of the matrix representation of 
Cartan algebra in the special unitary group $SU(3)$; namely, 
$\mu_{I} = (1,-1,0)(\mu_{\rm u},\mu_{\rm d},\mu_{\rm s})^{\rm t}$ and 
$\mu_{Y} = (1/2)(1,1,-2)(\mu_{\rm u},\mu_{\rm d},\mu_{\rm s})^{\rm t}$. 
Equation \eqref{chemical potential 2+1} gives 
\begin{eqnarray}
\renewcommand{\arraystretch}{1.2}
\begin{array}{l}
\mu_{\rm u} = \frac{1}{3}\mu_{B} + \frac{1}{2}\mu_{I} + \frac{1}{3}\mu_{Y},\\
\mu_{\rm d} = \frac{1}{3}\mu_{B} - \frac{1}{2}\mu_{I} + \frac{1}{3}\mu_{Y},\\
\mu_{\rm s} = \frac{1}{3}\mu_{B} - \frac{2}{3}\mu_{Y}.
\end{array}
\label{2+1chemical potential}
\end{eqnarray}
The coefficients on the right-hand side of Eq.~\eqref{2+1chemical potential}
correspond to the quantum numbers of u, d, s quarks. 
In this sense, the definition \eqref{chemical potential 2+1} is natural.

Also in the 2+1+1 flavor system, we can define the following 
relations by using Cartan algebra in the special unitary group $SU(4)$ 
for $\mu_{I}$, $\mu_{Y}$ and $\mu_{Y_c}$: 
\begin{eqnarray}
\renewcommand{\arraystretch}{1.2}
\begin{array}{l}
\mu_{B} = \frac{3}{4}(\mu_{\rm u} + \mu_{\rm d} + \mu_{\rm s}+ \mu_{\rm c}),\\
\mu_{I} = \mu_{\rm u} - \mu_{\rm d},\\
\mu_{Y} = \frac{1}{2}(\mu_{\rm u} + \mu_{\rm d} - 2\mu_{\rm s}),\\
\mu_{Y_c} = \frac{1}{3}(\mu_{\rm u} + \mu_{\rm d} + \mu_{\rm s} - 3\mu_{\rm c}),
\label{eq:mu 2+1+1}
\end{array}
\label{chemical potential 2+1+1}
\end{eqnarray}
where the quantum number $Y_c$ has been defined by 
$Y_c=(3/4)B-C$ with charmness $C$.  
Equation~\eqref{eq:mu 2+1+1} leads to 
\begin{eqnarray}
\renewcommand{\arraystretch}{1.2}
\begin{array}{l}
\mu_{\rm u} = \frac{1}{3}\mu_{B} + \frac{1}{2}\mu_{I} + \frac{1}{3}\mu_{Y} + \frac{1}{4}\mu_{Y_c},\\
\mu_{\rm d} = \frac{1}{3}\mu_{B} - \frac{1}{2}\mu_{I} + \frac{1}{3}\mu_{Y} + \frac{1}{4}\mu_{Y_c},\\
\mu_{\rm s} = \frac{1}{3}\mu_{B} - \frac{2}{3}\mu_{Y} + \frac{1}{4}\mu_{Y_c},\\
\mu_{\rm c} = \frac{1}{3}\mu_{B} - \frac{3}{4}\mu_{Y_c}.
\end{array}
\label{2+1+1chemical potential}
\end{eqnarray}
This final form is also natural, since the coefficients on 
the right-hand side of Eq.~\eqref{2+1+1chemical potential} are 
the quantum numbers of u, d, s, c quarks. 
Equation \eqref{chemical potential 2+1+1} is thus a natural extension of 
Eq.~\eqref{chemical potential 2+1}.

\subsection{Hadron resonance gas model}
\label{Hadron resonance gas model}
\vspace{-5pt}
For the hadron phase at low $T$, we use the HRG model. 
In the HRG model, the thermodynamic potential density ${\rm \Omega_{\rm H}}$ is described by free hadron gases. For convenience, ${\rm \Omega_{\rm H}}$ is 
divided into the baryonic piece ${\rm \Omega_{B}}$ and 
the mesonic one ${\rm \Omega _M}$:
\begin{eqnarray}
{\rm \Omega_{\rm H}}={\rm \Omega_{\rm B}}+{\rm \Omega_{\rm M}} 
\label{Omega_HRG}
\end{eqnarray}
with
\begin{eqnarray}
{\rm \Omega_{\rm{B}}} &=& -\sum_{i \in \rm Baryon}d_{{\rm B},i}T\int \frac{d^3{\bf p}}{(2\pi)^3} \big\{ \log(1+ e^{-(E_{{\rm B},i} - \mu_{{\rm B},i})/T})\nonumber\\
&&+\log(1+ e^{-(E_{{\rm B},i} + \mu_{{\rm B},i})/T})\big\}
\nonumber\\ 
\label{Omega_B}
\end{eqnarray}
and 
\begin{eqnarray}
{\rm \Omega_{\rm{M}}} &=& \sum_{j \in \rm Meson}d_{{\rm M},j}T\int \frac{d^3{\bf p}}{(2\pi)^3} \big\{ \log(1- e^{-(E_{\rm M,j}-\mu_{{\rm M},j})/T})\nonumber\\
&&+\log(1- e^{-(E_{{\rm M},j}+\mu_{{\rm M},j})/T})\big\}
\nonumber\\ 
\label{Omega_M}
\end{eqnarray}
for $E_{{\rm B},i}=\sqrt{{\bf p}^2+{m_{{\rm B},i}}^2}$ and $E_{{\rm M},j}=\sqrt{{\bf p}^2+{m_{{\rm M},j}}^2}$. Here, $m_{{\rm B},i}$ ($m_{{\rm M},j}$) and $\mu_{{\rm B},i}$ ($\mu_{{\rm M},j}$) is the mass and the chemical potential of the $i$-th baryon ($j$-th meson), respectively. 
In Eqs. \eqref{Omega_B} and \eqref{Omega_M}, all the hadrons listed in 
the Particle Data Table \cite{PDG} are taken; note that 
hadrons composed of u, d, s (u, d, s, c) quarks are picked up 
for the 2+1 (2+1+1) flavor system. 
The pressure $P_{\rm H}$ and the entropy density $s_{\rm H}$ are obtained 
from $\rm \Omega_{\rm H}$ as 
\begin{eqnarray}
P_{\rm H}&=&P_{\rm B}+P_{\rm M};~~~~~P_{\rm B}=-{\rm \Omega_{\rm B}},~~~~~P_{\rm M}=-{\rm \Omega_{\rm M}}, 
\label{P_H}
\\
s_{\rm H}&=&s_{\rm B}+s_{\rm M};~~~~~s_{\rm B}={\partial P_{\rm B}\over{\partial T}},~~~~~s_{\rm M}={\partial P_{\rm M}\over{\partial T}}. 
\label{s_H}
\end{eqnarray}

\begin{figure}[h]
\centering
\includegraphics[width=0.4\textwidth]{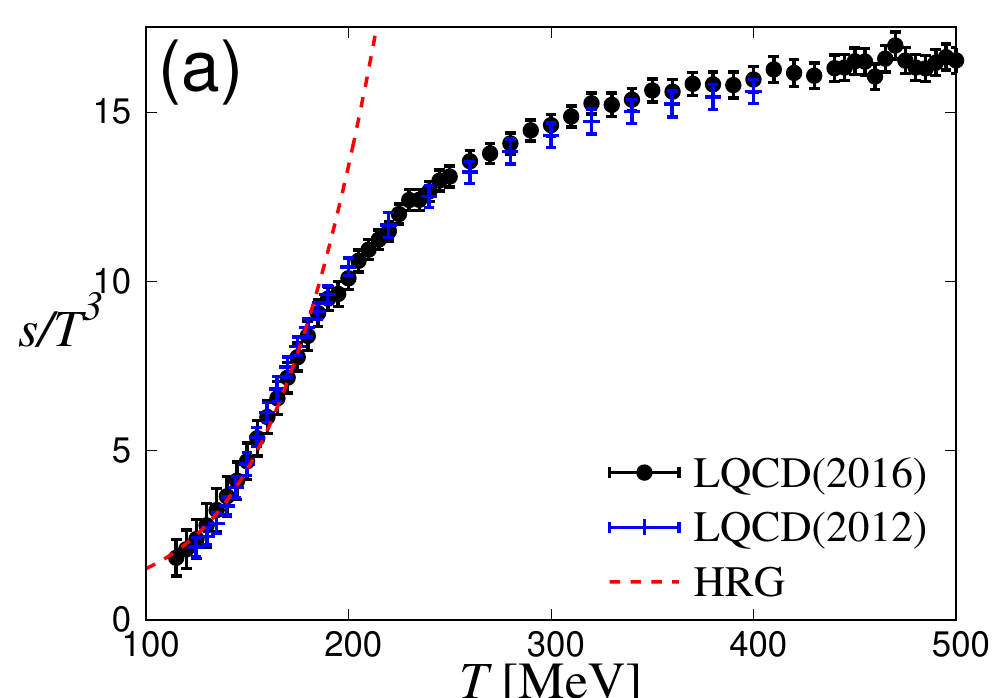}
\includegraphics[width=0.4\textwidth]{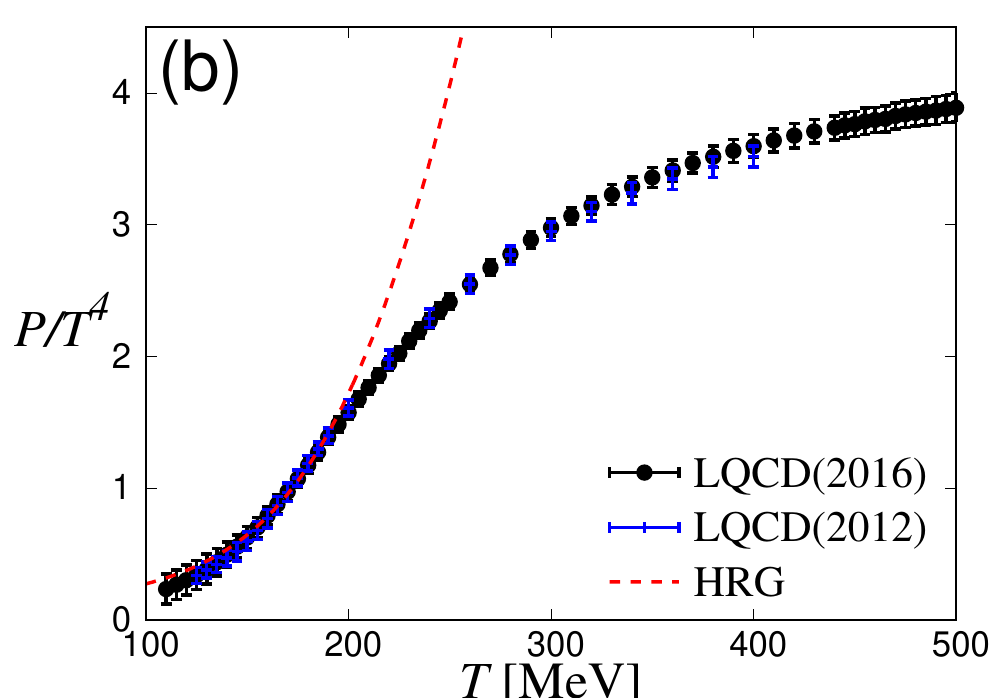}
\caption{$T$ dependence of (a) the entropy density $s$ and (b) 
the pressure $P$  
in the 2+1 flavor system with zero chemical potential. 
The dotted line means the result of the HRG model. 
LQCD data are taken from Refs.~\cite{Borsanyi_entropy,Borsanyi:2016ksw}.
In Ref.~\cite{Borsanyi:2016ksw}, LQCD data are available for $P$ 
but not for $s$. 
The entropy density $s$ is then evaluated by differentiating $P$ 
with respect to $T$. 
} 
\label{had P s 2+1}
\end{figure}

Figure~\ref{had P s 2+1} shows the entropy density $s$ and the pressure $P$ 
as a function of $T$ for the 2+1 flavor system with zero chemical potential. 
The HRG model (dotted line) well reproduces LQCD data \cite{Borsanyi_entropy,Borsanyi:2016ksw} in $T \lsim 170$~MeV for $s$ and $T \lsim 190$~MeV for $P$. 
The hadron phase is thus realized in $T \lsim 170$~MeV.

\begin{figure}[t]
\centering
\includegraphics[width=0.4\textwidth]{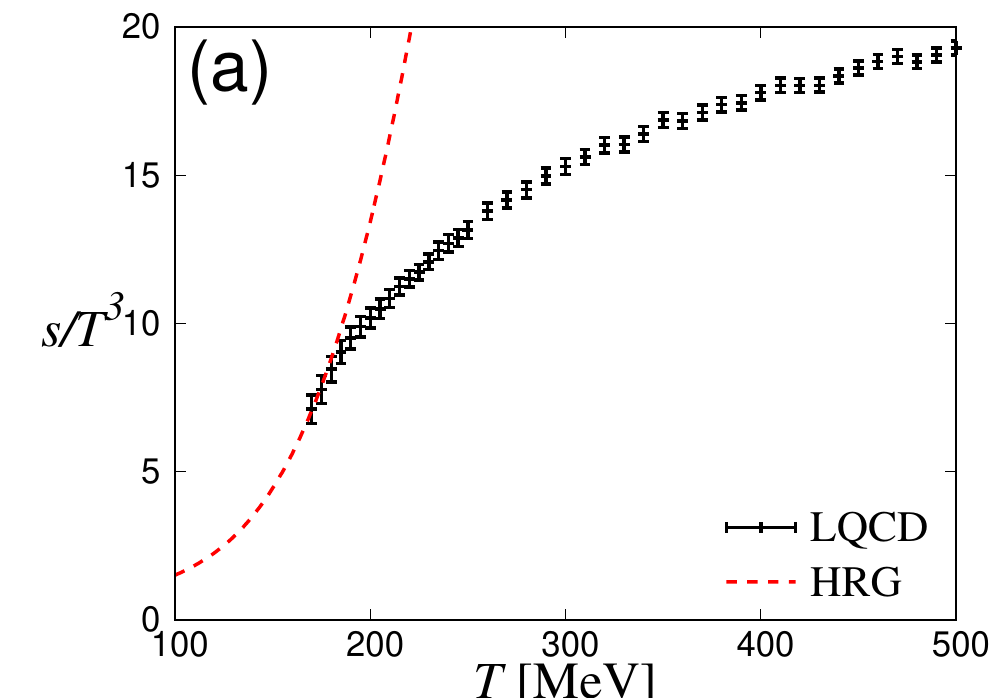}
\includegraphics[width=0.4\textwidth]{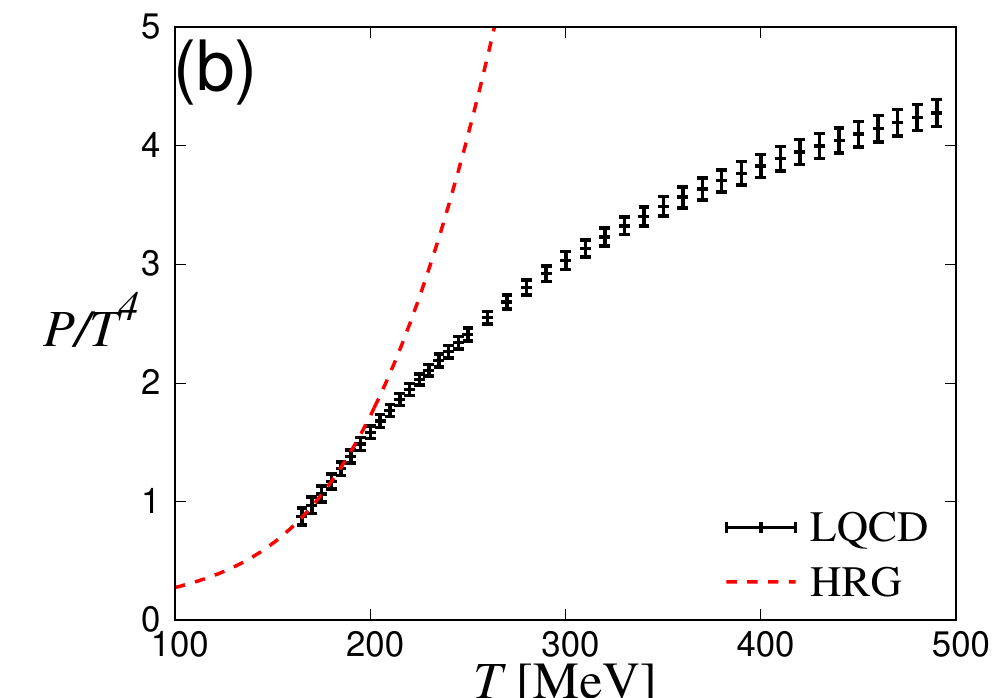}
\caption{$T$ dependence of (a) the entropy density $s$ and (b) 
the pressure $P$ 
in the 2+1+1 flavor system with zero chemical potential. 
The dotted line means the result of the HRG model. 
In Ref.~\cite{Borsanyi:2016ksw}, LQCD data are available for $P$ 
but not for $s$. 
The entropy density $s$ is then evaluated by differentiating $P$ 
with respect to $T$.}
\label{had P s 2+1+1}
\end{figure}

Figure~\ref{had P s 2+1+1} is the same as Fig.~\ref{had P s 2+1},  
but for 2+1+1 flavor system with zero chemical potential. 
The HRG model well explains LQCD data 
\cite{Borsanyi:2016ksw} 
in $T \lsim 190$~MeV for the pressure 
and $T \lsim 170$~MeV for the entropy density.
We find from Fig.~\ref{had P s 2+1+1} that the hadron phase is realized 
in $T \lsim 170$~MeV.

\subsection{Independent quark model}
\label{Independent quark model}
\begin{figure}[h]
\centering
\includegraphics[width=0.4\textwidth]{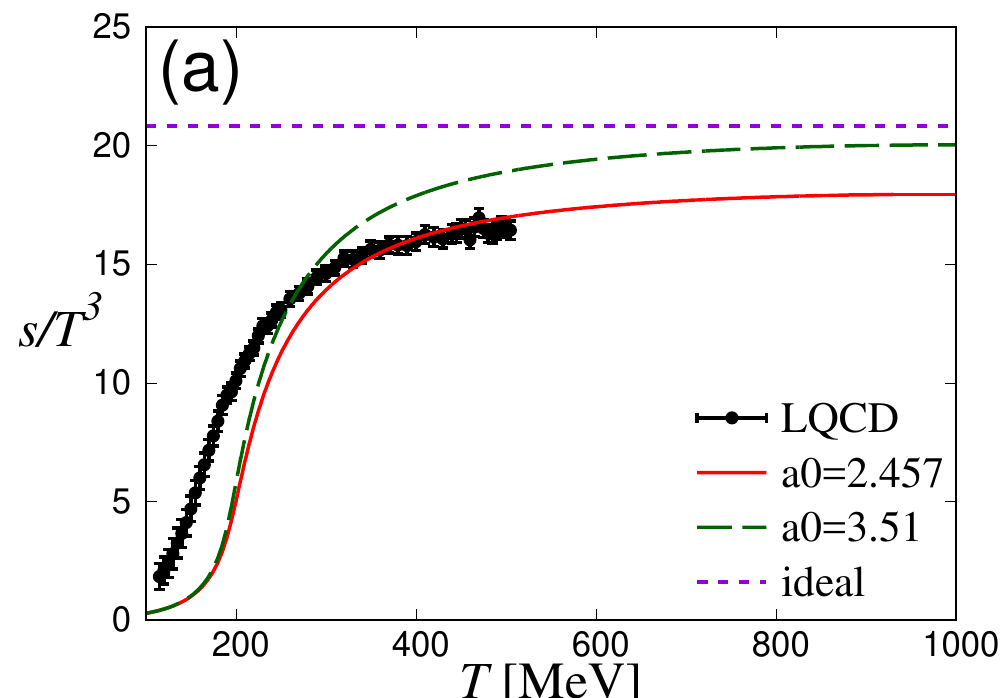}
\includegraphics[width=0.4\textwidth]{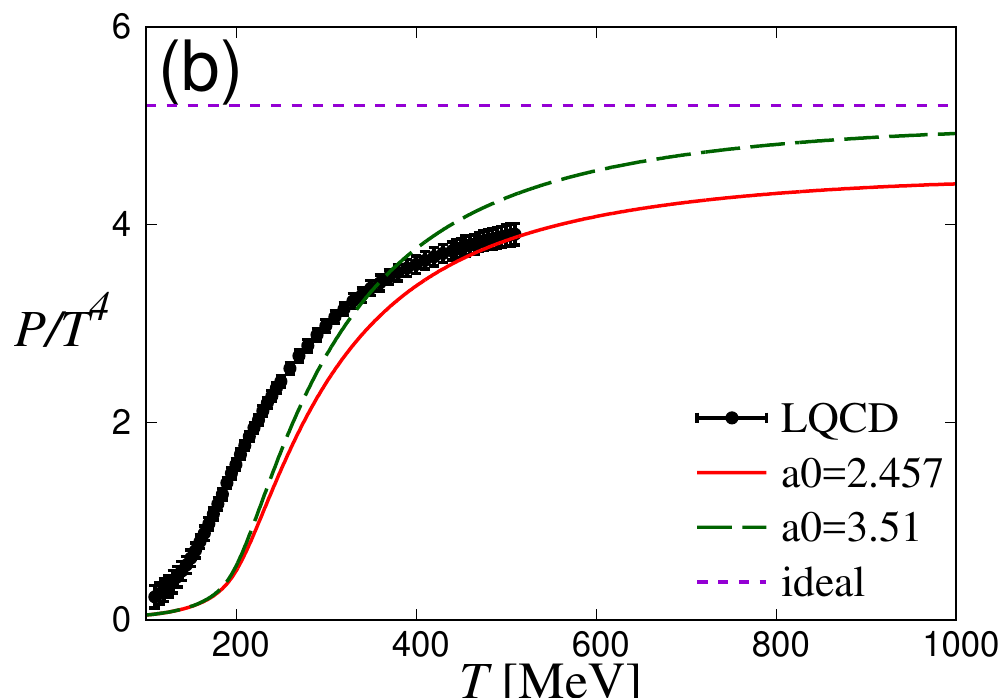}
\caption{$T$ dependence of (a) the entropy density $s$ and 
(b) the pressure $P$ 
for the 2+1 flavor system with zero chemical potential. 
The dotted line is the result of the ideal-gas model 
(the Stefan-Boltzmann limit), the dashed line denotes 
the IQ model with the original value $a_0=3.51$, and 
the solid line corresponds to the IQ model with $a_0=2.457$. 
LQCD data of Ref.~\cite{Borsanyi:2016ksw} 
are denoted by dots. 
}
\label{entropy_qg+2+1}
\end{figure}

\vspace{-5pt}
Next, we consider the quark phase that may appear 
in the region $T \approx 400$~MeV where 
LQCD data is consistent with NNLO HTL
perturbation~\cite{Haque:2013sja}. 
As shown later in Fig.~\ref{entropy_qg+2+1}, 
the entropy density calculated with LQCD simulations is about 80\% of 
the Stefan-Boltzmann limit value even at $T=500$~MeV. 
This means that the massless ideal-gas (massless-free-particle) model does not work. 
For this reason, we consider the PNJL model 
without any quark-quark direct interactions, 
since the interactions are expected to be 
not important above the hadron-quark transition temperature. 
In the model, quarks interact with the gluon field $A_{\mu}$ by 
the gauge coupling $g$, but the spatial parts $A_{i}$ $(i = 1,2,3)$ 
are neglected and 
only the temporal part $A_0$ is treated 
as a stationary and uniform background field. 
In this sense, we call this model ``independent quark (IQ) model".

As the gluonic action, we take the Polyakov-loop potential ${\cal U}$
 used in the PNJL model. The Lagrangian density of this model is 
\begin{eqnarray}
{\cal L}_{\rm Q}=\sum_f \left\{\bar{q}_f(i\gamma^{\mu}D_{\mu}-m_f )q_f \right\}-{\cal U}(T,{\Phi},{\bar{\Phi}}), 
\label{L_Q}
\end{eqnarray}
where $m_f$ is the current mass of $f$ quark 
and $D_{\mu}=\partial_{\mu}-igA_{\mu}^a\frac{\lambda_a}{2}\delta^{\mu0}$ with the Gell-Mann matrix $\lambda_a$ in color space. In Eq.~\eqref{L_Q}, the $q_f$ mean u, d, s 
quark fields for the 2+1 flavor system and u, d, s, c quark fields 
for the 2+1+1 flavor system.
See Ref.~\cite{Miyahara:2016din} for the definition of the Polyakov-loop ${\Phi}$ and its conjugate $\bar{\Phi}$.

Making the path integral over quark fields, the Lagrangian (\ref{L_Q}) yields 
the thermodynamic potential density,
\begin{eqnarray}
{\rm \Omega}_{\rm Q}&=&\ {\cal U}(T,{\Phi},{\bar{\Phi}}) 
\nonumber\\
&&- 2\sum_{f} \Bigg[ 
\int_{|{\bf p}|\leq \Lambda_{\rm T}} \frac{d^3{\bf p}}{(2\pi)^3} ( T\log{z_f^+}+ T\log{z_f^-})\Bigg]\nonumber\\
\label{Omega_Q}
\end{eqnarray}
for the quark matter.
The functions $z_f^+$ and $z_f^-$ are defined by 
\begin{eqnarray}
z_f^+ &=& 1+3{\bar{\Phi}} e^{-(E_f+\mu_f)/T}+3{\Phi} e^{-2(E_f+\mu_f)/T} 
\nonumber\\
&&+ e^{-3(E_f+\mu_f)/T},
\label{zfp}\\
z_f^- &=& 1+3{\Phi} e^{-(E_f-\mu_f)/T}+3{\bar{\Phi}} e^{-2(E_f-\mu_f)/T}\nonumber\\
&&+ e^{-3(E_f-\mu_f)/T}
\label{zfm}
\end{eqnarray}
with $E_f=\sqrt{{\bf p}^2+m_f^2}$. 
In Eq.~(\ref{Omega_Q}), the vacuum term has been omitted, since the pressure 
calculated with LQCD simulations does not include the term. 
The pressure $P_{\rm Q}$ and the entropy density $s_{\rm Q}$ are obtained from 
${\rm \Omega_{\rm Q}}$ as
\begin{eqnarray}
P_{\rm Q}&=&-{\rm \Omega_{\rm Q}}, 
\label{P_Q}
\\
s_{\rm Q}&=&{\partial P_{\rm Q}\over{\partial T}}. 
\label{s_Q}
\end{eqnarray}

We take the Polyakov-loop potential of Ref.~\cite{Rossner}:
\begin{eqnarray}
&&\frac{{\cal U}(T,{\Phi},{\bar{\Phi}})}{T^4}=-\frac{a(T)}{2}{\Phi}{\bar{\Phi}}\nonumber\\
&&+ b(T)\log\{1-6{\Phi}{\bar{\Phi}} + 4({\Phi}^3 + {\bar{\Phi}}^3) - 3({\Phi} {\bar{\Phi}})^2\};
\label{U_Phi}\\
&&a(T) = a_0 + a_1\left(\frac{T_0}{T}\right)+ a_2\left(\frac{T_0}{T}\right)^2,\\
&&b(T) = b_3\left(\frac{T_0}{T}\right)^3.   
\label{parameter_Phi}
\end{eqnarray}
In Ref.~\cite{Rossner}, the parameters $a_0$, $a_1$, $a_2$, $b_3$ and $T_0$ 
were fitted to LQCD data on the EoS in the pure gauge theory. 
In high $T$, the potential is dominated by the $a_0$ term. For this reason, 
the value of $a_0/2$ is set to the Stefan-Boltzmann limit value ($a_0=3.51$) in the pure gauge theory, but the $\cal{U}$ thus obtained overestimates new LQCD data~\cite{Borsanyi:2012pol} in 
$400 \lsim T \lsim 500$~MeV. 
In the 2+1 flavor system with dynamical quarks, furthermore, the quarks may 
change 
the Polyakov-loop potential. In fact, even at high $T$ such as $T=500$~MeV, 
the pressure $P$ calculated with LQCD simulations \cite{Borsanyi:2016ksw} is 
about 80~\% of the Stefan-Boltzmann limit value in 2+1 flavor system; 
see for example Fig.~\ref{entropy_qg+2+1}. 
The entropy density, obtained by differentiating $P$ with respect to $T$, also 
has the same property. This point is discussed below. 

In our previous work of Ref.~\cite{Miyahara:2016din} for the 
2+1 flavor system, the momentum cutoff $\Lambda_{\rm T}$ 
in the thermal quark-loop term of 
${\rm \Omega_Q}$ was determined 
to reproduce LQCD data \cite{Borsanyi_entropy} 
on $s$ at $T=300$ MeV, where the data were available 
in $T \leq 400$~MeV. 
However, we found that the model result 
with  the resulting value $\Lambda_{\rm T}=1.95$~GeV 
begins to underestimate new LQCD data 
\cite{Borsanyi:2016ksw} as $T$ increases from 400~MeV; 
here, we have evaluated 
the LQCD data on $s$ from new data \cite{Borsanyi:2016ksw} 
on $P$ measured in $T \leq 500$~MeV  by differentiating $P$ 
with respect to $T$. 
For this reason, we do not introduce $\Lambda_{\rm T}$ in this paper: Namely, 
$\Lambda_{\rm T} = \infty$.  
As mentioned above, in the 2+1 flavor system, 
the entropy density $s$ calculated with LQCD simulations underestimates 
 the Stefan-Boltzmann limit value by about 20~\% even at $T=500$~MeV. 
We then change the parameter $a_0$ so that the model result can reproduce 
the LQCD data at $T=400$~MeV. The resulting value is 
$a_0=0.7 \times 3.51=2.457$; 
see Table~\ref{Table_cal_U} for the values of parameters in ${\cal U}$. 
The model result with $a_0=2.457$
well explains LQCD data on $s$ in 
$400 \lsim T \leq 500$~MeV, as shown below. 
For the 2+1+1 flavor system, we keep $a_0=2.457$ to hold the simplicity of 
model.

\begin{table}[h]
\centering
\begin{tabular}{lcccr}
\hline
$\ a_0$&$a_1$&$a_2$&$b_3$&$T_0\hspace{5mm}$
\\ \hline
2.457 &-2.47&15.2&-1.75&270{\rm [MeV]}\\
\hline
\end{tabular}
\caption{Parameters in the Polyakov-loop potential. 
}
\label{Table_cal_U}
\end{table}

Figure~\ref{entropy_qg+2+1} shows 
the entropy density $s$ and the pressure $P$ 
as a function of $T$ 
for the 2+1 flavor system with zero chemical potential. 
LQCD data are taken from Ref.~\cite{Borsanyi:2016ksw} with large lattice.  
For the data, the entropy densities $s$ have been obtained 
by differentiating $P$ with respect to $T$. 
LQCD data are smaller than the result of the ideal-gas model 
(the Stefan-Boltzmann limit; dotted line) by about 20\% even at $T=500$~MeV. 
The IQ model with the original 
value $a_0=3.51$ (dashed line) underestimates 
the Stefan-Boltzmann limit by about 10~\% at $T=500$~MeV. 
Meanwhile, the model with $a_0=2.457$ (solid line) well explains LQCD data 
in the region $400 \lsim T \leq 500$~MeV. 
Thus, the quark phase may be realized in $T \gsim 400$ MeV. 
The lower limit of the quark phase is determinable 
clearly with $T$ dependence of $\chi_{ff'}^{(2)}$ ($f \neq f'$). This analysis 
will be made later in Sec. \ref{results}.

\begin{figure}[h]
\centering
\includegraphics[width=0.4\textwidth]{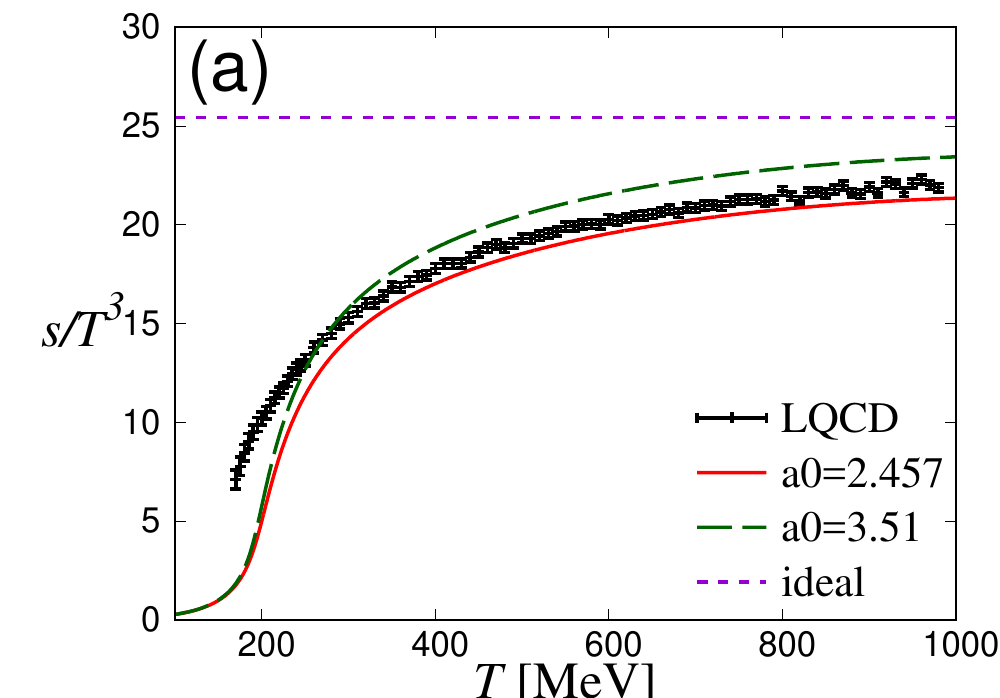}
\includegraphics[width=0.4\textwidth]{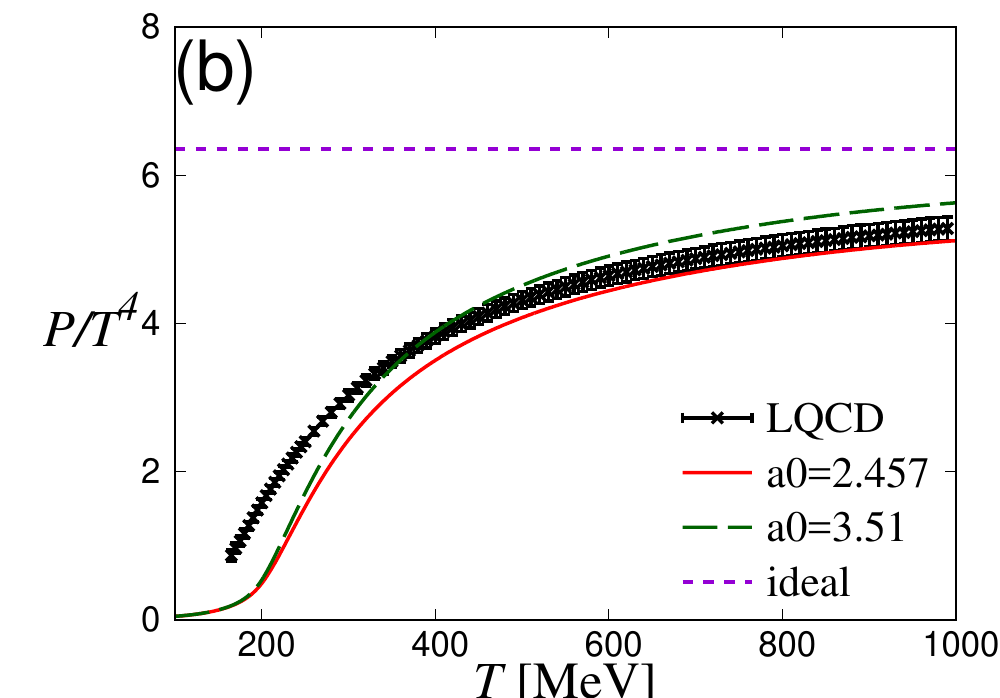}
\caption{$T$ dependence of (a) the entropy density $s$ and 
(b) the pressure $P$ for the 2+1+1 flavor system 
with zero chemical potential. 
See Fig.~\ref{entropy_qg+2+1} for the definition of lines. 
LQCD data (dots) are taken from Ref.~\cite{Borsanyi:2016ksw}; 
see the text for further explanation. 
}
\label{entropy_qg-2+1+1}
\end{figure}

Figure~\ref{entropy_qg-2+1+1} shows the entropy density $s$ and 
the pressure $P$ as a function of $T$ 
for the 2+1+1 flavor system with zero chemical potential. 
LQCD calculations were done for $P$ in Ref.~\cite{Borsanyi:2016ksw}. 
The entropy density $s$ is evaluated from the data 
by differentiating $P$ with respect to $T$. 
LQCD data are about 80\% of the result of the ideal-gas model 
(the Stefan-Boltzmann limit; dotted line) at $T=1000$~MeV. 
The IQ model with the original 
value $a_0=3.51$ (dashed line) 
reaches about 90\% of the Stefan-Boltzmann limit value at $T=1000$~MeV. 
The model with $a_0=2.457$ (solid line) reproduces LQCD data 
in $400 \lsim T \leq 1000$~MeV pretty well. 
Thus, the quark phase may be realized in $T \gsim 400$~MeV 
also for the 2+1+1 flavor system. 
The lower limit of the quark phase can be determined 
precisely with $T$ dependence of $\chi_{ff'}^{(2)}$ ($f \neq f'$). 
This analysis is also made in Sec. \ref{results}.

\subsection{Quark-hadron crossover model}
Now we consider the HQC model defined by Eq.~\eqref{s_Hybrid_def} in which $s_{\rm H}(T,\{\mu_\a \})$ is calculated by  
the HRG model of Sec.~\ref{Hadron resonance gas model} 
and $s_{\rm Q}(T,\{\mu_\a\})$ is by 
the IQ model of Sec.~\ref{Independent quark model}, 
where $\{\mu_\a\}$ means 
$\{\mu_{B},\mu_{I},\mu_{Y}\}$ for the 2+1 flavor system and 
$\{\mu_{B},\mu_{I},\mu_{Y}, \mu_{\rm Y_c}\}$ for the 2+1+1 
flavor system. 
This model is a natural extension of the model of  
Ref.~\cite{Hatsuda_hybrid model} in which quarks and gluons are treated as 
ideal gases. 
The function $f_{\rm H}(T,\{\mu_\a\})$ means 
the hadron-production probability.  
When $f_{\rm H}(T,\{\mu_\a \})=1~(0)$, the system consists 
of hadron matter (quark gluon matter) only, i.e., the system becomes 
the mixed phase in $0<f_{\rm H}(T,\{\mu_\a \})<1$. 

In this paper, we consider $P(T)$ and $s(T)$ as the EoS and 
several kinds of second-order susceptibilities 
for the 2+1 and 2+1+1 flavor systems with zero chemical potential.

\subsubsection{2+1 flavor system}
\label{II-2+1 flavor system}
\vspace{-10pt}
We recapitulate the formalism of Ref.~\cite{Miyahara:2016din}. 
For the 2+1 flavor system, the 
$f_{\rm H}(T,\{\mu_\a \})$ is expanded 
into  a power series of $\{\mu_\a\}$ and is taken up to the second order:
\begin{eqnarray}
f_{\rm H}(T,\{\mu_{\a}\}) &=& f^{(0)}_{\rm H}(T) + f_{{\rm H},B}^{(2)}(T)\left(\frac{\mu_{B}}{T_{\rm c}}\right)^2\nonumber\\
&&+ f_{{\rm H},I}^{(2)}(T)\left(\frac{\mu_{I}}{T_{\rm c}}\right)^2 + f_{{\rm H},Y}^{(2)}(T)\left(\frac{\mu_{Y}}{T_{\rm c}}\right)^2\nonumber\\
&&+ f_{{\rm H},BY}^{(2)}(T)\left(\frac{\mu_{B}}{T_{\rm c}}\right)\left(\frac{\mu_{Y}}{T_{\rm c}}\right).\nonumber\\
\label{Eq:F_H}
\end{eqnarray}
where $T_{\rm c}=170$~MeV \cite{Borsanyi_Tc,YAoki_Tc^(Z_3)} 
is the hadron-quark transition 
temperature defined with the Polyakov loop. 
The form of Eq.~\eqref{Eq:F_H} comes from two properties; 
(i) $s$ is invariant under charge conjugation, i.e., 
the transformation $(\mu_{B},\mu_{I},\mu_{Y})\rightarrow(-\mu_{B},-\mu_{I},-\mu_{Y})$, 
and (ii) the system is also invariant under the interchange 
$\mu_{\rm u} \leftrightarrow \mu_{\rm d}$, i.e., 
the transformation $(\mu_{B},\mu_{I},\mu_{Y})\rightarrow(\mu_{B},-\mu_{I},\mu_{Y})$. 
For $\mu_{B}=\mu_{I}=\mu_{Y}=0$, 
Eq.~\eqref{s_Hybrid_def} reduces to 
\begin{eqnarray}
s(T)=f_{\rm H}^{(0)}(T) s_{\rm H}(T)+\{1-f_{\rm H}^{(0)}(T)\} s_{\rm Q}(T). 
\label{s_Hybrid_0}
\end{eqnarray}

The pressure $P$ with no vacuum contribution is obtainable from 
$s$ as 
\begin{eqnarray}
&&P(T,\{\mu_{\a}\})\nonumber\\
&=&\int_{0}^{T}dT' s(T',\{\mu_{\a}\})
\nonumber\\
&=& 
\int_{0}^{T}dT' s_{\rm Q} 
+ \int_{0}^{T}dT' f_{\rm H}\left(s_{\rm H}-s_{\rm Q}\right)
\nonumber\\
&=&P_{{\rm Q}}(T,\{\mu_{\a}\}) 
+ \int_{0}^{T}dT' f_{\rm H}\left(s_{\rm H}-s_{\rm Q}\right) .
\label{P_Hybrid}
\end{eqnarray}
The second-order diagonal susceptibility $\chi^{(2)}_{\alpha}$ 
of quantum numbers $\a=B, I, Y$  
is obtained as the second derivative of $P$ 
with respect to the chemical potential $\mu_{\alpha}$: 
\begin{eqnarray}
&&\chi^{(2)}_{\alpha}(T,\{\mu_{\a}\}))\nonumber\\
&=&\frac{\partial^2}{\partial \mu_{\alpha}^2} P(T,\{\mu_{\a}\})\nonumber\\
&=& \frac{\partial^2}{\partial \mu_{\alpha}^2} P_{\rm Q}(T,\{\mu_{\a}\}) 
+ \int_0^T dT' \Bigg[\frac{\partial^2 f_{\rm H}}{\partial \mu_{\alpha}^2}(s_{\rm H}-s_{\rm Q})\nonumber\\
&&+2\frac{\partial f_{\rm H}}{\partial \mu_{\a}}\frac{\partial (s_{\rm H}-s_{\rm Q})}{\partial \mu_{\alpha}}+ f_{\rm H}\frac{\partial^2 (s_{\rm H}-s_{\rm Q})}{\partial \mu_{\alpha}^2}\Bigg] .
\label{chi_a} 
\end{eqnarray}
Particularly at $\mu_{B}=\mu_{I}=\mu_{Y}=0$, it becomes  
\begin{eqnarray}
&&\chi^{(2)}_{\alpha}(T)\nonumber\\
&=& \frac{\partial^2}{\partial \mu_{\alpha}^2} P_{\rm Q}(T,\{\mu_{\a}\})|_{{\{\mu_\a \}}=0}
\nonumber\\ 
&&
+ \int_0^T dT' \left[ 2 f_{{\rm H},\a}^{(2)}(s_{\rm H}-s_{\rm Q})
+ f_{\rm H}^{(0)} 
\frac{\partial^2 (s_{\rm H}-s_{\rm Q})}{\partial \mu_{\alpha}^2}\right].\nonumber\\ 
\label{chi_a_mu=0}
\end{eqnarray}
Similarly, the $BY$ correlation susceptibility is 
\begin{eqnarray}
&&\chi^{(2)}_{BY}(T,\{\mu_{\a}\})\nonumber\\
&=&\frac{\partial^2}{\partial \mu_{B}\partial \mu_{Y}} 
P(T,\{\mu_{\a}\})
\nonumber\\
&=& \frac{\partial^2}{\partial \mu_{B}\partial \mu_{Y}} 
P_{\rm Q}(T,\{\mu_{\a}\})  + \int_0^T dT' \Bigg[\frac{\partial^2 f_{\rm H}}{\partial \mu_{B}\partial \mu_{Y}}(s_{\rm H}-s_{\rm Q})\nonumber\\
&&+\frac{\partial f_{\rm H}}{\partial \mu_{B}}\frac{\partial (s_{\rm H}-s_{\rm Q})}{\partial \mu_{Y}}+\frac{\partial f_{\rm H}}{\partial \mu_{Y}}\frac{\partial (s_{\rm H}-s_{\rm Q})}{\partial \mu_{B}}
\nonumber\\
&&+ f_{\rm H}\frac{\partial^2 (s_{\rm H}-s_{\rm Q})}{\partial \mu_{B}\partial \mu_{Y}}\Bigg] 
\label{chi_by}
\end{eqnarray}
for finite $\{\mu_{\a}\}$ and    
\begin{eqnarray}
&&
\chi^{(2)}_{BY}(T)\nonumber\\
&=& \frac{\partial^2}{\partial \mu_{B}\partial \mu_{Y}} 
P_{\rm Q}(T,\{\mu_{\a}\})|_{{\{\mu_\a \}}=0} \nonumber\\
&& + \int_0^T dT' \left[f_{{\rm H},BY}^{(2)} (s_{\rm H}-s_{\rm Q})
+ f_{\rm H}^{(0)}\frac{\partial^2 (s_{\rm H}-s_{\rm Q})}{\partial \mu_{B}\partial \mu_{Y}}\right] \nonumber\\
\label{chi_by_mu=0}
\end{eqnarray}
for $\{\mu_{\a}\}=0$.

Using Eqs.~\eqref{s_Hybrid_0},~\eqref{chi_a_mu=0},~\eqref{chi_by_mu=0}, 
one can determine $f_{\rm H}^{(0)}$, $f_{{\rm H},{\a}}^{(2)}$, $f_{{{\rm H},BY}}^{(2)}$ 
from LQCD data on $s$, $\chi^{(2)}_{\alpha}$, $\chi^{(2)}_{BY}$ 
at $\{\mu_{\a}\}=0$, respectively: Namely, 
\begin{eqnarray} 
f_{\rm H}^{(0)}
&=& \frac{s^{\rm LQCD} - s_{\rm Q}}{s_{\rm H} - s_{\rm Q}}
\label{def fH0}
\end{eqnarray}
and 
\begin{eqnarray} 
f_{{\rm H},\gamma}^{(2)}
&=& \frac{1}{w(s_{\rm H} - s_{\rm Q})}\Bigg[\frac{\partial \chi_{\gamma}^{(2),{\rm LQCD}}}{\partial T} - \frac{\partial \chi_{\gamma}^{(2),{\rm Q}}}{\partial T}\nonumber\\
&&- f_{\rm H}^{(0)}\left(\frac{\partial \chi_{\gamma}^{(2),{\rm H}}}{\partial T} - \frac{\partial \chi_{\gamma}^{(2),{\rm Q}}}{\partial T} \right)\Bigg] 
\label{diagonal fH2}
\end{eqnarray}
for $\gamma=\a, BY$, where 
the superscript ``LQCD" means LQCD data, 
$w=2$ for $\gamma=\a$ and $w=1$ for  $\gamma = BY$, and 
\bea
\chi_{\a}^{(2),{\rm Q}}=
\left.
\frac{ \partial^2 P_{\rm Q}}{\partial \mu_{\a}^2}
\right|_{{\{\mu_\a \}}=0}
,~~~~ 
\chi_{BY}^{(2),{\rm Q}}=
\left.
\frac{ \partial^2 P_{\rm Q}
}{\partial \mu_{B}\partial \mu_{Y} } 
\right|_{{\{\mu_\a \}}=0}.~~~
\eea

\begin{figure}[H]
\centering
\vspace{0cm}
\includegraphics[width=0.4\textwidth]{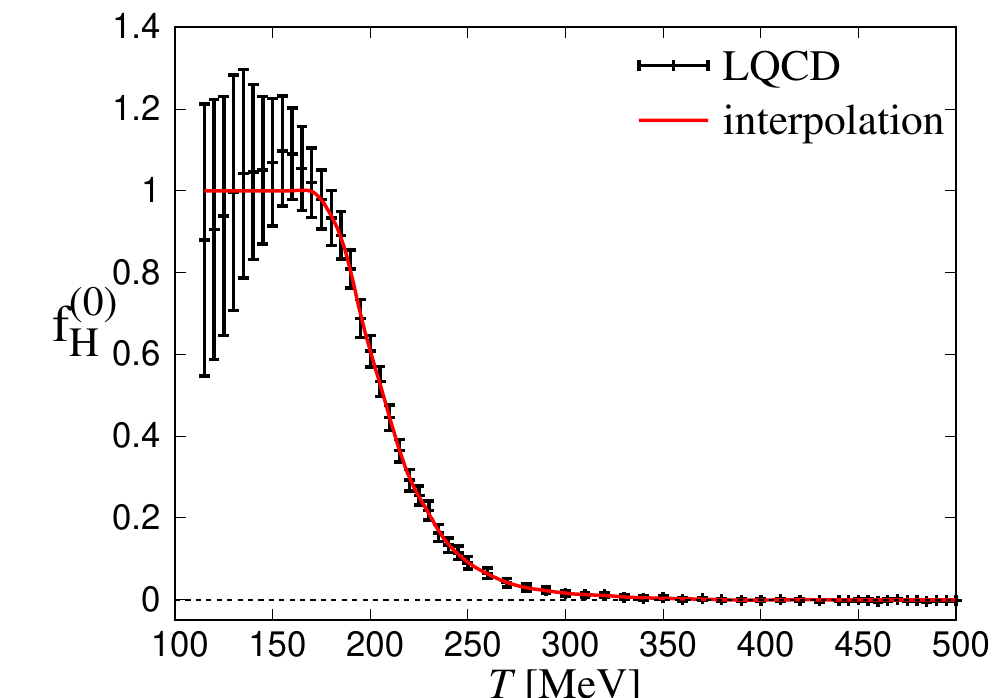}
\caption{$T$ dependence of $f_{\rm H}^{(0)}(T)$.
The solid line is the smooth function obtained with 
the cubic spline interpolation. 
LQCD data on $f_{\rm H}^{(0)}$ (dots) are deduced 
from those~\cite{Borsanyi:2016ksw} on $s$ 
by using Eq.~\eqref{s_Hybrid_0}. 
}
\label{fig_f_0_v2}
\end{figure}

Figure \ref{fig_f_0_v2} shows the $f_{\rm H}^{(0)}$ (dots with error bars) 
deduced from 
LQCD data~\cite{Borsanyi:2016ksw} on $s$ by using Eq.~\eqref{def fH0}. 
We make the cubic spline interpolation for the mean values of data in 
order to obtain the smooth function that passes through the mean values. 
Here, the mean values have been taken in $175 \le T \le 400$~MeV 
where the mean values are smaller than 1, and have been set to 0 in $T > 400$~MeV 
where the mean values are quite small. 
In $T \le 170$~MeV, LQCD data have large error 
bars and the mean values are not so reliable; 
in fact, the mean values are accidentally larger than 1 in 
$140< T \le 170$~MeV. 
For this reason, we have replaced the mean values by 1 in $T \le 150$~MeV and 
have neglected the mean values in $150 < T \le 170$~MeV.   
The smooth function thus obtained (solid line) is consistent 
with LQCD data; note that the function is very close to 1 
in $150 < T \le 170$~MeV.
Figure \ref{fig_f_0_v2} indicates that 
the mixed phase appears in a region $170 \lsim T \lsim 400$~MeV for the case 
of zero chemical potential.  
Here, it should be noted 
that in Eq.~\eqref{s_Hybrid_def} the hadron piece $f_{\rm H}s_{\rm H}$ 
contributes to $s$ up to 400~MeV since $s_{\rm H}$ increases rapidly as $T$ 
increases. The upper limit of the phase transition can be determined 
clearly from the off-diagonal susceptibilities. 
This will be shown later in Sec. \ref{III:2+1 system}.

\begin{figure}[H]
\centering
\includegraphics[width=0.4\textwidth]{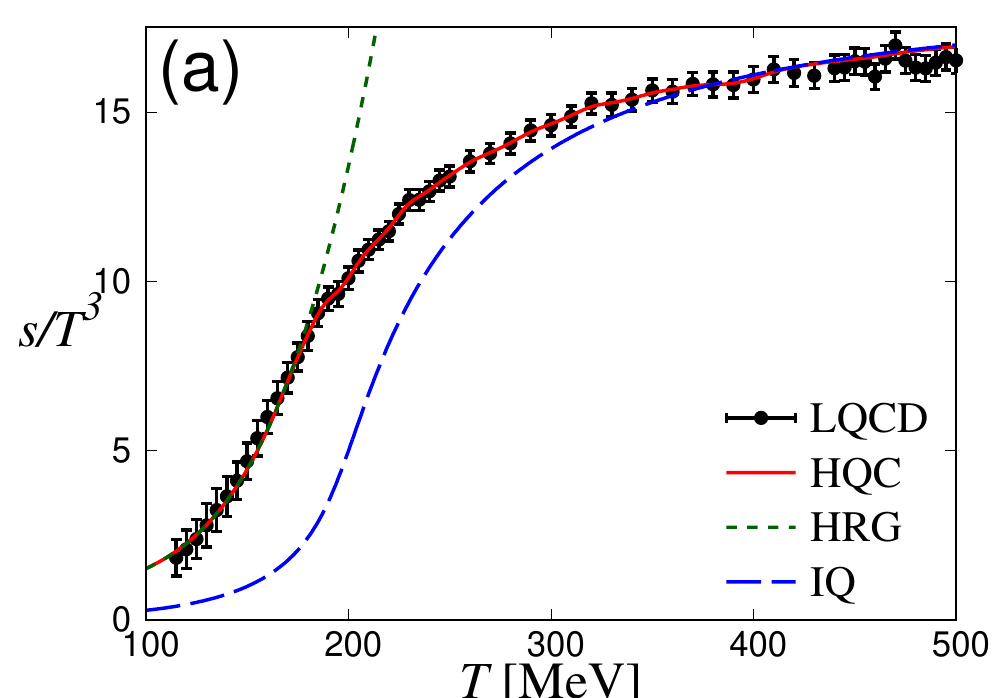}
\includegraphics[width=0.4\textwidth]{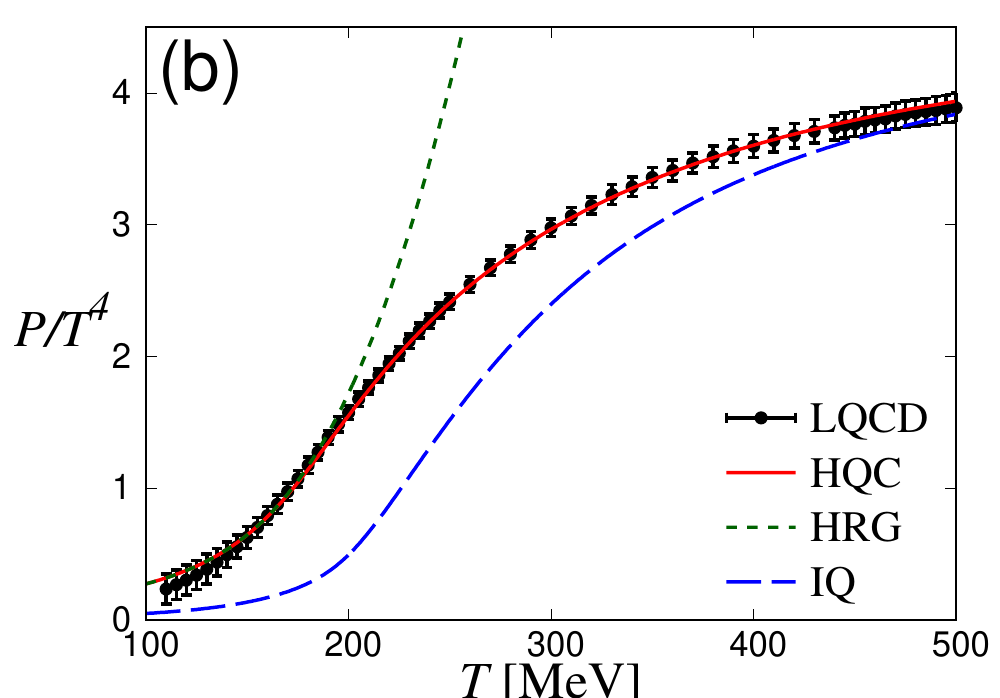}
\caption{$T$ dependence of the entropy density $s$ and the pressure 
$P$ calculated by the HQC model for the 2+1 flavor system with 
zero chemical potential. 
The solid line is the HQC result based on the 
$f_{\rm H}^{(0)}(T)$ determined in Fig.~\ref{fig_f_0_v2}. 
The dotted line stands for the result of the HRG model, 
the dashed line corresponds to that of the IQ model. 
In Ref.~\cite{Borsanyi:2016ksw}, 
LQCD data are available for $P$  but not for $s$. 
The entropy density $s$ is then evaluated by differentiating $P$ 
with respect to $T$.
}
\label{Determination of f-0_v2}
\end{figure}

Figure \ref{Determination of f-0_v2} shows $s(T)$ and 
$P(T)$ as a function of $T$ in the case of zero chemical potential. 
The HQC model with the $f_{\rm H}^{(0)}(T)$ 
determined above should reproduce LQCD data automatically. 
This is satisfied, as shown by the solid line.

\begin{figure}[h]
\centering
\hspace{0mm}
\includegraphics[width=0.4\textwidth]{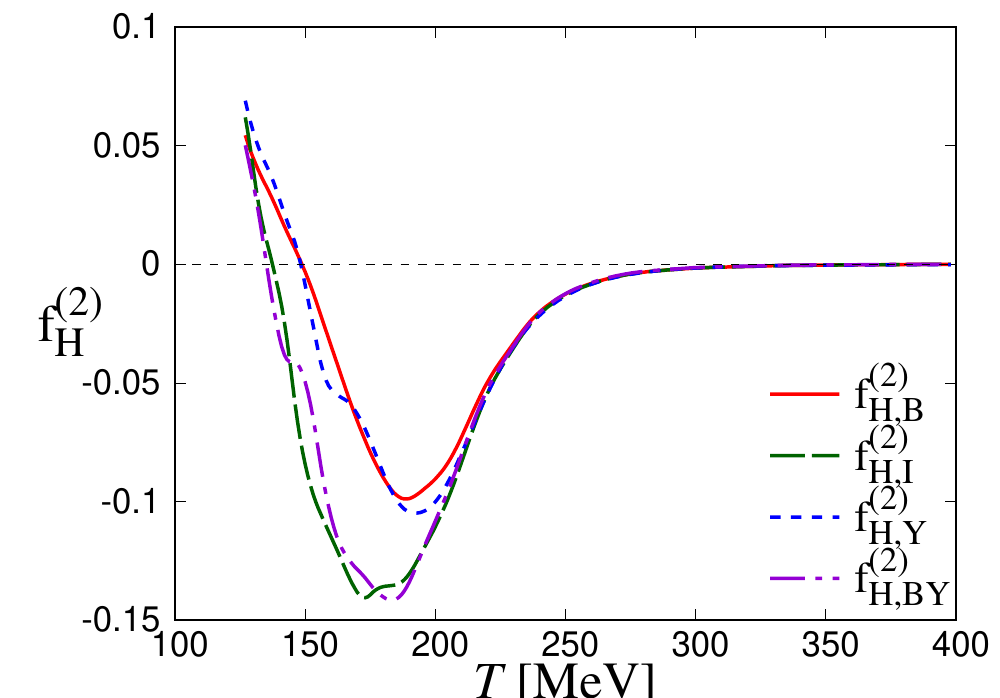}
\caption{
Results of the cubic spline interpolation for 
$T$ dependence of $f_{{\rm H},{\gamma}}^{(2)}$ 
that are deduced from LQCD data~\cite{Borsanyi:2016ksw} on 
$\chi^{(2)}_{\gamma}$ by using Eq.~\eqref{diagonal fH2}. 
The results are plotted by the solid line for $f_{{\rm H},{B}}^{(2)}$, 
the dashed line for $f_{{\rm H},{I}}^{(2)}$, 
the dotted line for $f_{{\rm H},{Y}}^{(2)}$, 
and the dot-dashed line for $f_{{\rm H},{BY}}^{(2)}$. 
}
\label{fig_f_gamma_v2}
\end{figure}

\begin{figure*}[t]
\begin{tabular}{cc}
\begin{minipage}{0.5\hsize}
\centering
\includegraphics[width=0.79\textwidth]{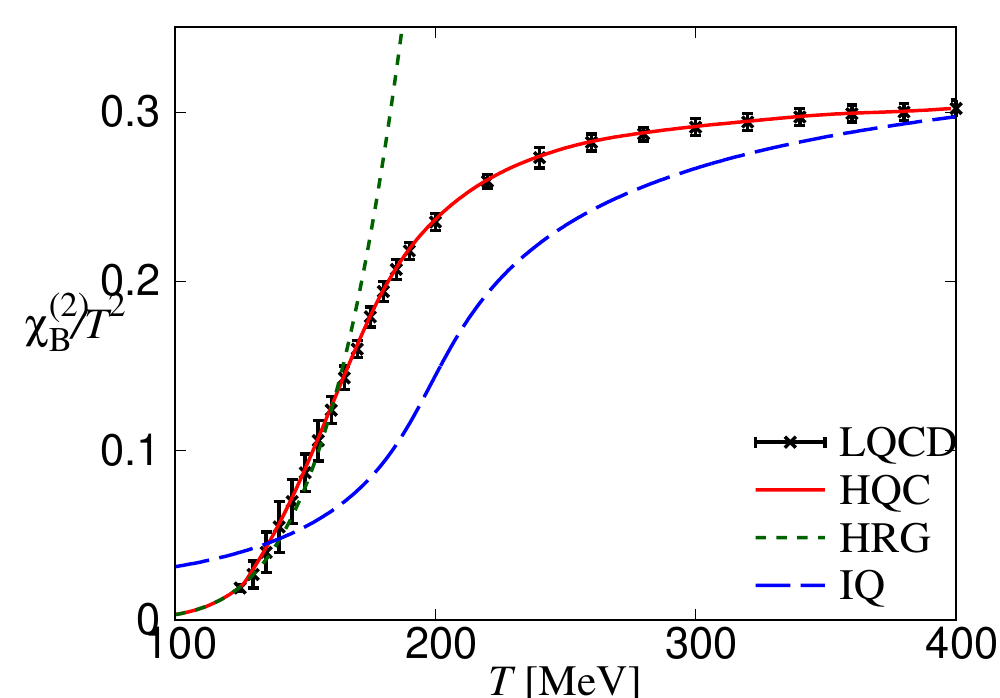}
\end{minipage}
\begin{minipage}{0.5\hsize}
\centering
\includegraphics[width=0.79\textwidth]{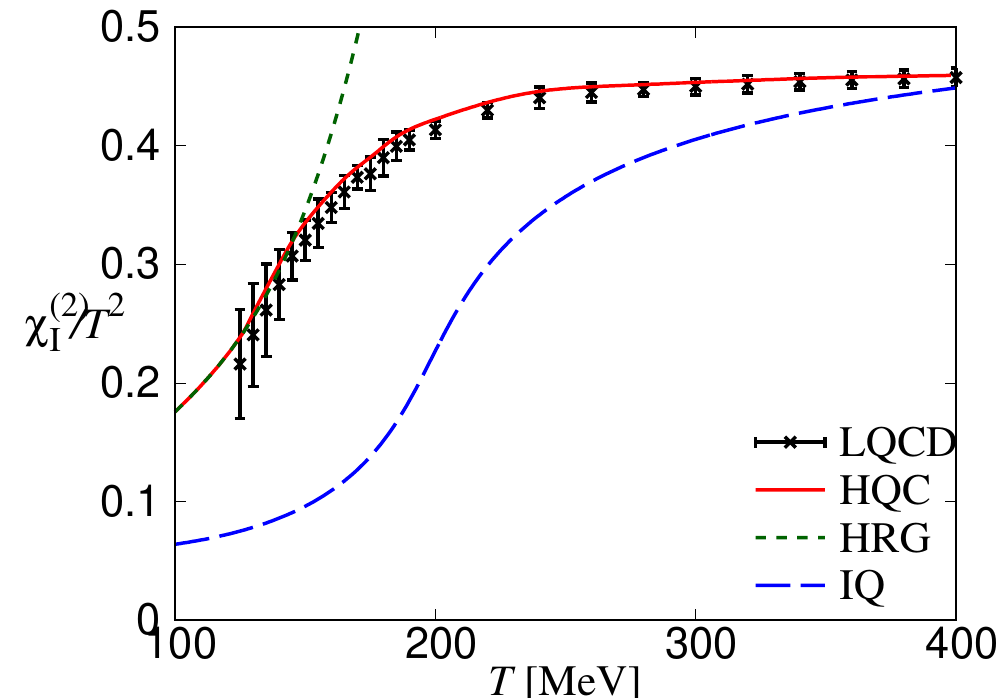}
\end{minipage}\\
\begin{minipage}{0.5\hsize}
\centering
\includegraphics[width=0.79\textwidth]{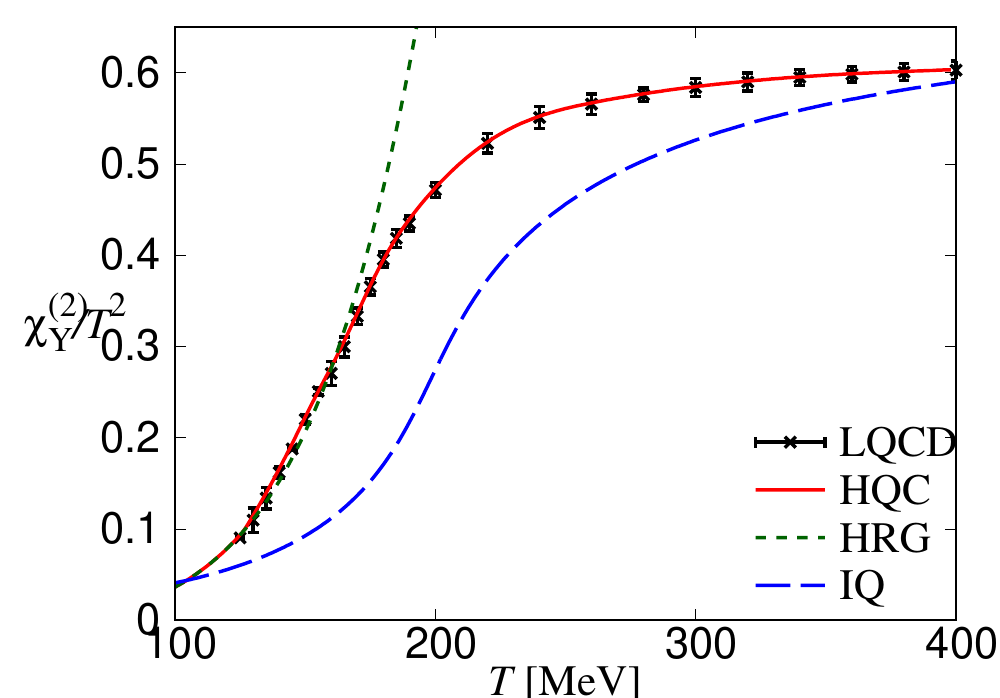}
\end{minipage}
\begin{minipage}{0.5\hsize}
\centering
\includegraphics[width=0.79\textwidth]{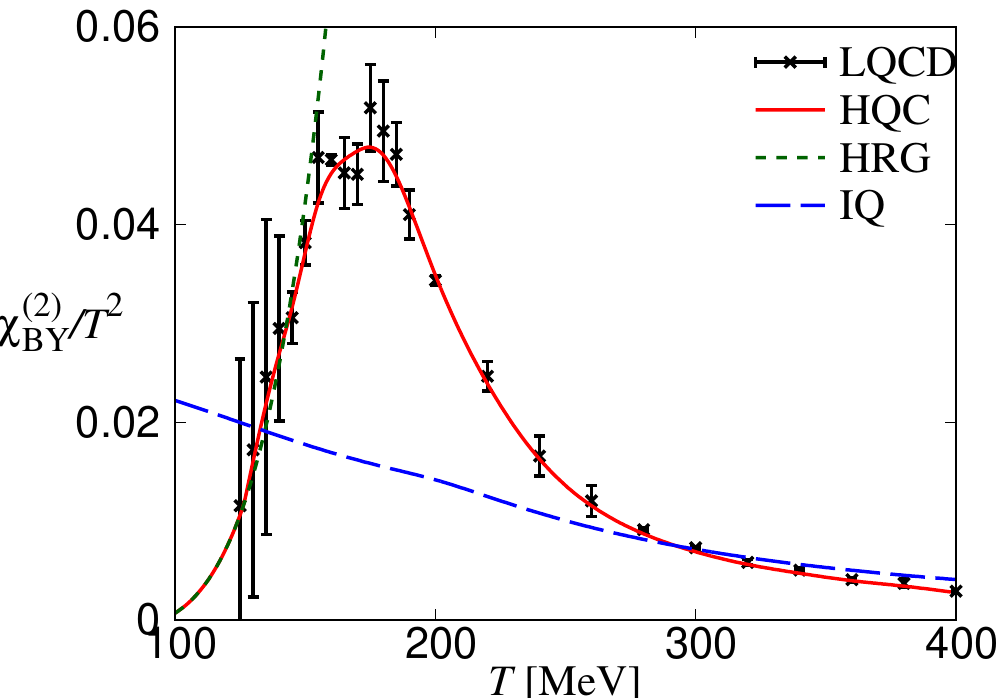}
\end{minipage}
\end{tabular}
\caption{$T$ dependence of the baryon-number $B$ and the isospin $I$, 
the hypercharge $Y$ and the $BY$ correlation susceptibility 
for the 2+1 flavor system with zero chemical potential.  
The HQC result is drawn by the solid line. 
The dotted line stands for the result of the HRG model, 
the dashed line corresponds to that of the IQ model. 
LQCD date (dots) are taken from Ref~\cite{Borsanyi:2016ksw}. 
}
\label{input susceptibility_v2}
\end{figure*}

We take the same procedure for $f_{{\rm H},{\gamma}}^{(2)}(T)$,  
where $\gamma=B, I, Y, BY$. Namely, the $f_{{\rm H},{\gamma}}^{(2)}(T)$ 
are deduced from LQCD data~\cite{Borsanyi:2016ksw} on 
$\chi^{(2)}_{\gamma}(T)$ by using Eq.~\eqref{diagonal fH2}, 
and the cubic spline interpolation is made for 
the mean values of the $f_{{\rm H},{\gamma}}^{(2)}(T)$. 
Here we have simply assumed $f_{{\rm H},{\gamma}}^{(2)}(T)=0$ 
in $T \le 127$~MeV where LQCD data are not available.  
The resulting smooth lines are plotted in Fig.~\ref{fig_f_gamma_v2}. 
All the $f_{{\rm H},{\gamma}}^{(2)}(T)$ are order ${\cal O}(10^{-1})$. 
In order to confirm the accuracy of the cubic spline interpolation, 
we compare the original LQCD data on 
$\chi^{(2)}_{\gamma}(T)$ with the corresponding HQC result (solid line)
in Fig.~\ref{input susceptibility_v2}. 
As expected, good agreement is seen between them.  
Again, the HRG model (dotted line) reproduces the LQCD data in 
$T \lsim 170$~MeV, while the IQ model well explains the data around 
$T = 400$~MeV. 

One can see from Fig.~\ref{fig_f_gamma_v2} that 
the $f_{{\rm H},{\gamma}}^{(2)}(T)$ satisfy 
\bea
f_{{\rm H},{B}}^{(2)}(T) \approx f_{{\rm H},{Y}}^{(2)}(T) \approx 
\frac{3}{4}f_{{\rm H},{I}}^{(2)}(T) \approx \frac{3}{4} 
f_{{\rm H},{BY}}^{(2)}(T) 
\label{equiv-BIY}
\eea
around $T=200$~MeV. The $f_{{\rm H},{\gamma}}^{(2)}(T)$ are thus close 
to each other around $T=200$~MeV. This property plays an important role 
when we draw the QCD phase diagram in 
$\mu_{B}$--$T$, $\mu_{I}$--$T$ $\mu_{Y}$--$T$ planes. 
This will be discussed later in Sec. \ref{III:2+1 system}

Throughout the analyses mentioned above, we have succeeded in determining 
$f_{\rm H}(T,\{\mu_\a \})$. The $f_{\rm H}$ can be described by 
$\mu_f$ ($f$=u, d, s) by using Eq.~\eqref{chemical potential 2+1}:
\bea
&&f_{\rm H}(T,\{\mu_\a \})\nonumber\\
&=& 
f^{(0)}_{\rm H}(T) + \sum_{f,f' \in {\rm u,d,s}} 
f_{{\rm H},ff'}^{(2)}(T)\left(\frac{\mu_{f}}{T_{\rm c}} \right)
\left(\frac{\mu_{f'}}{T_{\rm c}} \right)~~
\label{F_ff 2+1}
\eea
with
\bea
f_{\rm H,uu}^{(2)}(T) &=& f_{{\rm H},B}^{(2)}(T) + f_{{\rm H},I}^{(2)}(T) 
\nonumber\\
&&+ \frac{1}{4}f_{{\rm H},Y}^{(2)}(T) + \frac{1}{2}f_{{\rm H},BY}^{(2)}(T), 
\label{BIY-ff'-1}
\\
f_{\rm H,ss}^{(2)}(T) &=& f_{{\rm H},B}^{(2)}(T) + f_{{\rm H},Y}^{(2)}(T) - f_{{\rm H},BY}^{(2)}(T),
\label{BIY-ff'-2}
\\
f_{\rm H,ud}^{(2)}(T) &=& 2f_{{\rm H},B}^{(2)}(T) - 2f_{{\rm H},I}^{(2)}(T) 
\nonumber\\
&&+ \frac{1}{2}f_{{\rm H},Y}^{(2)}(T) + f_{{\rm H},BY}^{(2)}(T),
\label{BIY-ff'-3}
\\
f_{\rm H,us}^{(2)}(T) &=& 2f_{{\rm H},B}^{(2)}(T) - f_{{\rm H},Y}^{(2)}(T) - \frac{1}{2}f_{{\rm H},BY}^{(2)}(T); 
\label{BIY-ff'-4}
\eea
note that 
$f_{\rm H,uu}^{(2)}(T)=f_{\rm H,dd}^{(2)}(T), 
f_{\rm H,ud}^{(2)}(T)=f_{\rm H,du}^{(2)}(T)$ 
and $f_{\rm H,us}^{(2)}(T)=f_{\rm H,ds}^{(2)}(T)$. 

\begin{figure}[H]
\centering
\vspace{0cm}
\includegraphics[width=0.4\textwidth]{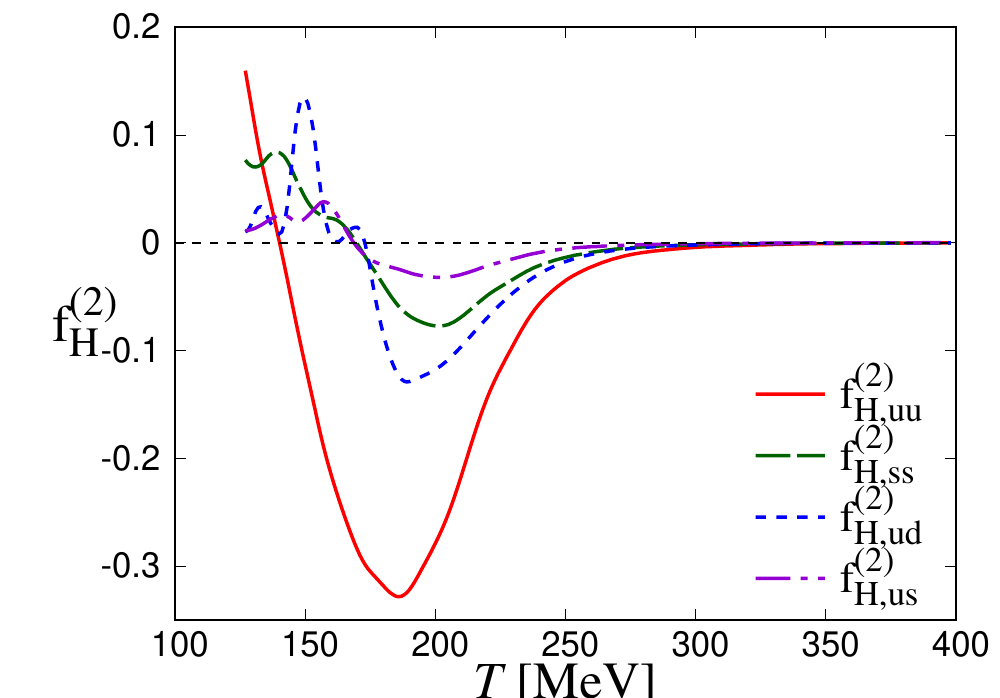}
\caption{
$T$ dependence of $f_{{\rm H},ff'}^{(2)}$ 
derived from  $f_{{\rm H},{\gamma}}^{(2)}$. 
The results are plotted by the solid line for $f_{\rm H,uu}^{(2)}$, 
the dashed line for $f_{\rm H,ss}^{(2)}$, 
the dotted line for $f_{\rm H,ud}^{(2)}$, 
and the dot-dashed line for $f_{\rm H,us}^{(2)}$. 
}
\label{fig_f_ff'}
\end{figure}

Figure \ref{fig_f_ff'} shows $T$ dependence of the $f_{{\rm H},ff'}^{(2)}$ 
that are derived from  the $f_{{\rm H},{\gamma}}^{(2)}$ 
by using Eqs. \eqref{BIY-ff'-1}--\eqref{BIY-ff'-4}. 
We can see from this figure that  
\bea
f_{\rm H,uu}^{(2)}(T) \approx 2.5 f_{\rm H,ud}^{(2)}(T) \approx 
4 f_{\rm H,ss}^{(2)}(T) \approx 10 f_{\rm H,us}^{(2)}(T) 
\label{eq:f_ff'}
\eea
around $T=200$~MeV. 
Thus, the s-quark contribution is small in 
the $f_{{\rm H},ff'}^{(2)}$ and also in the $f_{{\rm H},{\gamma}}^{(2)}$.
This property induces the approximate relation 
$f_{{\rm H},{B}}^{(2)}(T) \approx f_{{\rm H},{Y}}^{(2)}(T)$ shown in 
Eq.~\eqref{equiv-BIY}, since 
\bea
f_{{\rm H},B}^{(2)}(T) - f_{{\rm H},Y}^{(2)}(T) = \frac{1}{9}\Big[-3f_{\rm H,ss}^{(2)}(T) + 6f_{\rm H,us}^{(2)}(T)\Big] \approx 0 .
\nonumber\\
\label{relation-BY-ff}
\eea

The flavor diagonal and off-diagonal 
susceptibilities $\chi_{ff'}^{(2)}(T,\{\mu_{\a}\})$ 
are obtained from $P$ of 
Eq.~\eqref{P_Hybrid} by using Eq.~\eqref{F_ff 2+1} as $f_{\rm H}$: 
\bea
&&\chi_{ff'}^{(2)}(T,\{\mu_{\a}\}) 
\nonumber\\
&=& 
\frac{\partial^2}{\partial \mu_{f}\partial \mu_{f'}} 
P_{\rm Q}(T,\{\mu_{\a}\}) 
\nonumber\\
&&+ \int_0^T dT' \Bigg[\frac{\partial^2 f_{\rm H}}{\partial \mu_{f}\partial \mu_{f'}}(s_{\rm H}-s_{\rm Q}) + \frac{\partial f_{\rm H}}{\partial \mu_{f}}\frac{\partial (s_{\rm H}-s_{\rm Q})}{\partial \mu_{f'}} \nonumber\\
&&+\frac{\partial f_{\rm H}}{\partial \mu_{f'}}\frac{\partial (s_{\rm H}-s_{\rm Q})}{\partial \mu_{f}}+ f_{\rm H}\frac{\partial^2 (s_{\rm H}-s_{\rm Q})}{\partial \mu_{f}\partial \mu_{f'}}\Bigg] 
\label{flavor susceptibilities}
\eea
for finite $\{\mu_\a \}$ and 
\bea
&&\chi_{ff'}^{(2)}(T) 
\nonumber\\
&=& 
\chi_{ff'}^{(2),{\rm Q}}(T,\{\mu_{\a}\})|_{\{\mu_{\a}\}=0} 
\nonumber\\
&&+ \int_0^T dT' \Bigg[wf^{(2)}_{{\rm H},ff'}(s_{\rm H}-s_{\rm Q}) + 
f^{(0)}_{\rm H}\frac{\partial^2 (s_{\rm H}-s_{\rm Q})}{\partial \mu_{f}\partial \mu_{f'}}\Bigg]  \notag \\
\label{flavor susceptibilities-0}
\eea
for $\{\mu_\a \}=0$,
where $w=2$ for $f=f'$  and $1$ for $f \neq f'$. 
{It is known that 
the off-diagonal flavor susceptibilities 
$\chi_{ff'}^{(2),{\rm Q}}(T)$ 
of the PNJL-type model are negligibly small~\cite{Kouno_3f_FTBC}.  
Hence, for simplicity of calculation, we put 
$\chi_{ff'}^{(2),{\rm Q}}(T) =0$ for $f \neq f'$. }

\subsubsection{2+1+1 flavor system}
\vspace{-10pt}
Also for the 2+1+1 flavor system, the hadron-production probability can be described by 
\bea
&&f_{\rm H}^{2+1+1}(T,\{\mu_\a \})\nonumber\\
&&  = f^{2+1+1,(0)}_{\rm H}(T)\nonumber\\
&~~~~~~&\hspace{10pt} + \sum_{f,f' \in {\rm u,d,s,c}} 
f_{{\rm H},ff'}^{2+1+1,(2)}(T)\left(\frac{\mu_{f}}{T_{\rm c}} \right)
\left(\frac{\mu_{f'}}{T_{\rm c}} \right).~~~~~~~~
\label{F_ff 2+1+1}
\eea
In order to keep the simplicity of our model, however, we assume that 
the $f_{\rm H}$ of Eq.~\eqref{F_ff 2+1} is applicable for the 2+1 flavor subsystem of the 
2+1+1 flavor system, and that $f_{{\rm H,c}f}^{2+1+1,(2)} = 0$ for $f =$ u, d, s, c.
In this case, the HQC model has no adjustable parameter for 
the 2+1+1 flavor system. 
This assumption is justified later in Sec. \ref{III:2+1+1 system}. 
The procedure for obtaining the EoS and the flavor diagonal and off-diagonal 
susceptibilities is the same as in the 2+1 flavor system. Hereafter, we neglect the superscript ``2+1+1" in $f_{{\rm H}}^{2+1+1}$, when it does not induce any confusion.

\vspace{-10pt}
\section{Numerical results}
\label{results}
\vspace{-10pt}

\subsection{2+1 flavor system}
\label{III:2+1 system}
\begin{figure}[h]
\centering
\vspace{0cm}
\includegraphics[width=0.4\textwidth]{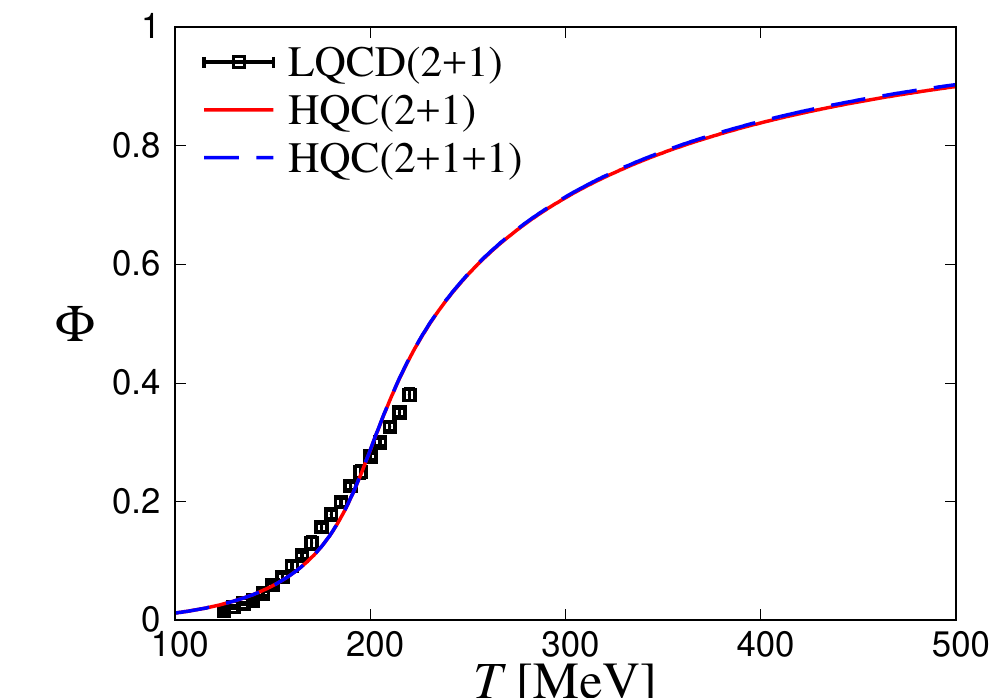}
\caption{$T$ dependence of the Polyakov loop $\Phi$
in the 2+1 and 2+1+1 flavor systems  with zero chemical potential. 
LQCD data are available for the 2+1 flavor system~\cite{Borsanyi_Tc,YAoki_Tc^(Z_3)}. 
The solid and dashed lines are results of our model 
for the 2+1 and 2+1+1 flavor systems, respectively. 
}
\label{fig_Poyakov-loop 2+1}
\end{figure}

\vspace{-10pt}
In general, the pseudocritical temperature $T_c^{(\cal{O})}$ of hadron-quark 
(confinement-deconfinement) crossover 
depends on observable $\cal{O}$ considered. 
The definition commonly used is the peak in the $T$ 
derivative of the Polyakov loop $\Phi$. 
Figure~\ref{fig_Poyakov-loop 2+1} shows $T$ dependence of 
the Polyakov loop $\Phi$. 
In the HQC model, $\Phi$ is calculated by the IQ model. 
LQCD data are available for the 2+1 flavor system~\cite{Borsanyi_Tc,YAoki_Tc^(Z_3)}. 
Our model (solid line) reproduces the LQCD data pretty well. 
The pseudocritical temperature $T_c^{(\Phi)}=201$~MeV of our model is somewhat 
larger than LQCD result $T_c^{(\Phi),{\rm LQCD}}=170 \pm 7$~MeV. 
The model result (dashed line) 
for the 2+1+1 flavor system is very close to that (solid line) 
for the 2+1 flavor system. This indicates that c quark hardly affects the 
hadron-quark transition. In practice, this supports 
that the $f_{\rm H}(T,\{\mu_{\a}\})$ determined from the 2+1 flavor LQCD data is applicable for the 2+1+1 flavor system, 
if we do not mind $\chi^{(2)}_{{\rm H,c}f}$ ($f=$u, d, s, c).

\vspace{-10pt}

As an alternative to $T_c^{(\Phi)}$, we may consider 
$f_{\rm H}(T,\{\mu_{\a}\})=1/2$ as a definition of $T_c$. 
Our result is $T_c^{(f_{\rm H})}=207$~MeV for $\{\mu_{\a}\}=0$ and 
somewhat larger than the LQCD result $T_c^{(\Phi)}=170 \pm 7$~MeV~\cite{Borsanyi_Tc,YAoki_Tc^(Z_3)}.

\begin{figure}[h]
\centering
\vspace{0cm}
\includegraphics[width=0.4\textwidth]{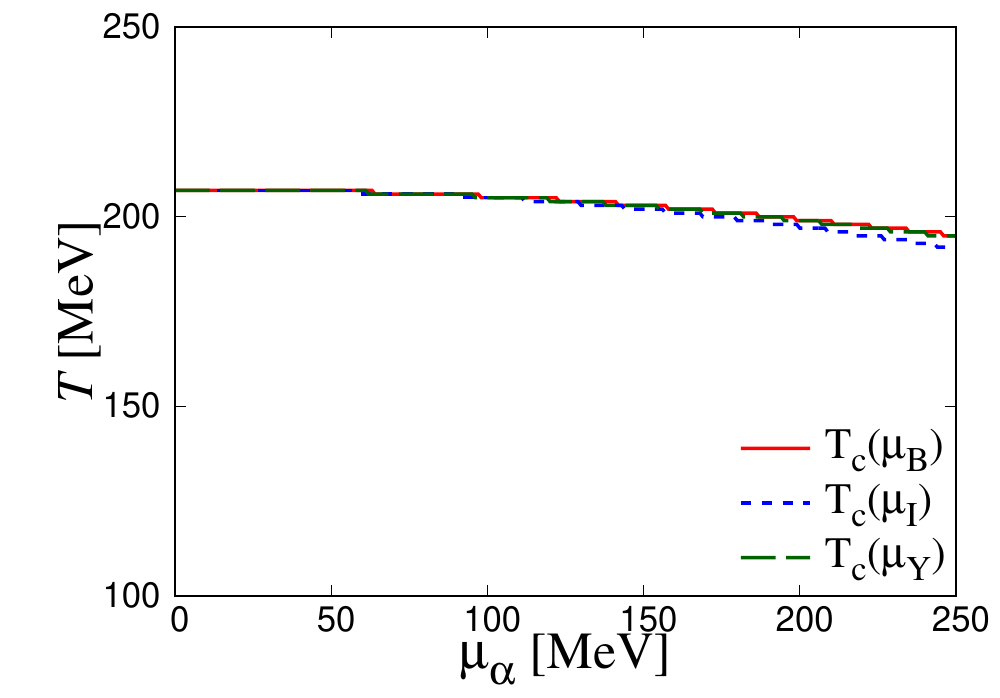}
\caption{Phase diagram in $\mu_{B}$--$T$, $\mu_{I}$--$T$, $\mu_{Y}$--$T$ planes. }
\label{fig_phase-diagram-BIY}
\end{figure}

Figure \ref{fig_phase-diagram-BIY} shows the QCD phase diagram 
in $\mu_{B}$--$T$, $\mu_{I}$--$T$, $\mu_{Y}$--$T$ planes. 
The symbol $T_c(\mu_{\a})$ denotes the pseudocritical temperature 
of the hadron-quark transition in $\mu_{\a}$--$T$ plane, 
where the pseudocritical temperature is defined by 
$f_{\rm H}=1/2$. 
In virtue of Eq.~\eqref{equiv-BIY}, the three transition lines 
almost agree with each other. Thus, 
the relation  
\bea
T_c(\mu_{B}) \approx T_c(\mu_{I}) \approx T_c(\mu_{Y}) 
\label{eq:BIY-equivalence}
\eea
is satisfied in $\mu_{\a} < 250$~MeV. 
We call this relation ``$BIY$ approximate equivalence" in the present paper. 
As for the $\mu_{B}$ direction, we can evaluate the fourth-order term 
$f_{{\rm H},B}^{(4)}$ from LQCD data on $\chi_{B}^{(4)}$. 
We have confirmed that the $f_{{\rm H},B}^{(4)}$ does not affect 
$T_c(\mu_{B})$ in $\mu_{B} < 250$~MeV.

$BIY$ approximate equivalence comes from the property of 
Eq.~\eqref{equiv-BIY}. As already mentioned in Sec.~\ref{II-2+1 flavor system}, the approximate relation 
$f_{{\rm H},B}^{(2)}(T) \approx f_{{\rm H},Y}^{(2)}(T)$ in 
Eq.~\eqref{equiv-BIY} comes from the fact that 
the s-quark contribution is small in the $f_{{\rm H},\gamma}^{(2)}(T)$ ($\gamma=B, I, Y, BY$). 
Meanwhile, the relation $f_{{\rm H},B}^{(2)}(T) \approx 
3 f_{{\rm H},Y}^{(2)}(T)/4$ in Eq.~\eqref{equiv-BIY}
is a remnant of the fact 
that  in the 2 flavor system $T_c(\mu_{B})=T_c(\mu_{I})$ when 
these are expressed 
up to $(\mu_{\a}/T)^2$ ($\a=B,I$)~\cite{Ejiri:2007ga}.

\begin{figure}[h]
\centering
\vspace{0cm}
\includegraphics[width=0.4\textwidth]{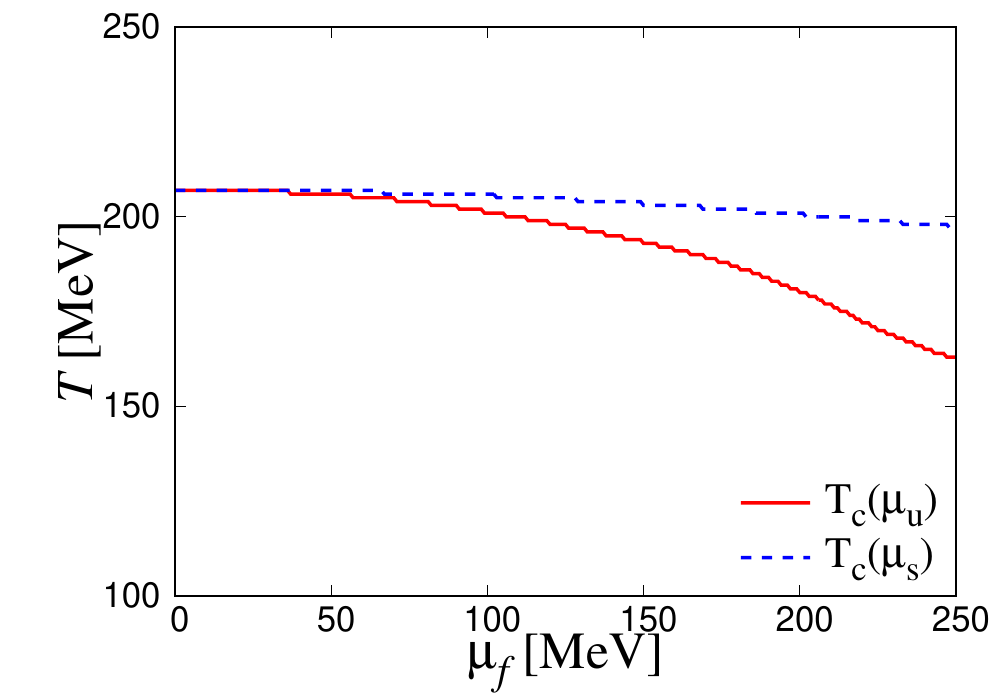}
\vspace{-10pt}
\caption{Phase diagram in $\mu_{u}$--$T$ and $\mu_{s}$--$T$ planes.
Note that $T_c(\mu_{\rm u})=T_c(\mu_{\rm d})$ $\mu_{\rm u}=\mu_{\rm d}$ 
because of $f_{\rm H,{uu}}^{(2)}(T)=f_{\rm H,{dd}}^{(2)}(T)$ and $f_{\rm H,{us}}^{(2)}(T)=f_{\rm H,{ds}}^{(2)}(T)$. 
}
\label{phase diagram f}
\end{figure}

Next, we consider the flavor dependence of the pseudocritical temperature of 
the hadron-quark transition.  
Figure \ref{phase diagram f} shows the QCD phase diagram 
in $\mu_{f}$--$T$ planes. 
The symbol $T_c(\mu_{f})$ denotes the pseudocritical temperature 
in $\mu_{f}$--$T$ plane. 
Note that $T_c(\mu_{\rm u})=T_c(\mu_{\rm d})$ for $\mu_{\rm u}=\mu_{\rm d}$, because 
of $f_{\rm H,{uu}}^{(2)}(T)=f_{\rm H,{dd}}^{(2)}(T)$ and $f_{\rm H,{us}}^{(2)}(T)=f_{\rm H,{ds}}^{(2)}(T)$.
We then draw $T_c(\mu_{\rm u})$ and $T_c(\mu_{\rm s})$ only 
in Fig.~\ref{phase diagram f}.
The transition takes place at higher $T$ in $\mu_{\rm s}$--$T$ plane than 
in $\mu_{\rm u}$--$T$ plane.  This may stem from the fact that $m_{\rm s} \gg 
m_{\rm u}=m_{\rm d}$. Further discussion will be made in 
Sec.~\ref{III:2+1+1 system}.

\vspace{-10pt}

\begin{figure*}[t]
\begin{tabular}{cc}
\begin{minipage}{0.5\hsize}
\centering
\includegraphics[width=0.79\textwidth]{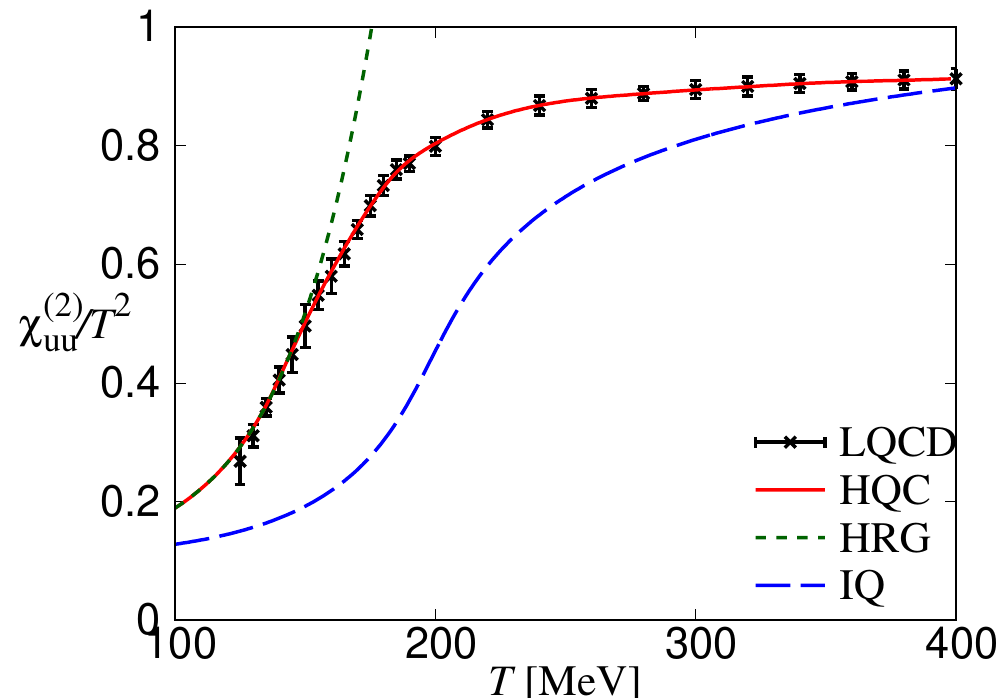}
\end{minipage}
\begin{minipage}{0.5\hsize}
\centering
\includegraphics[width=0.79\textwidth]{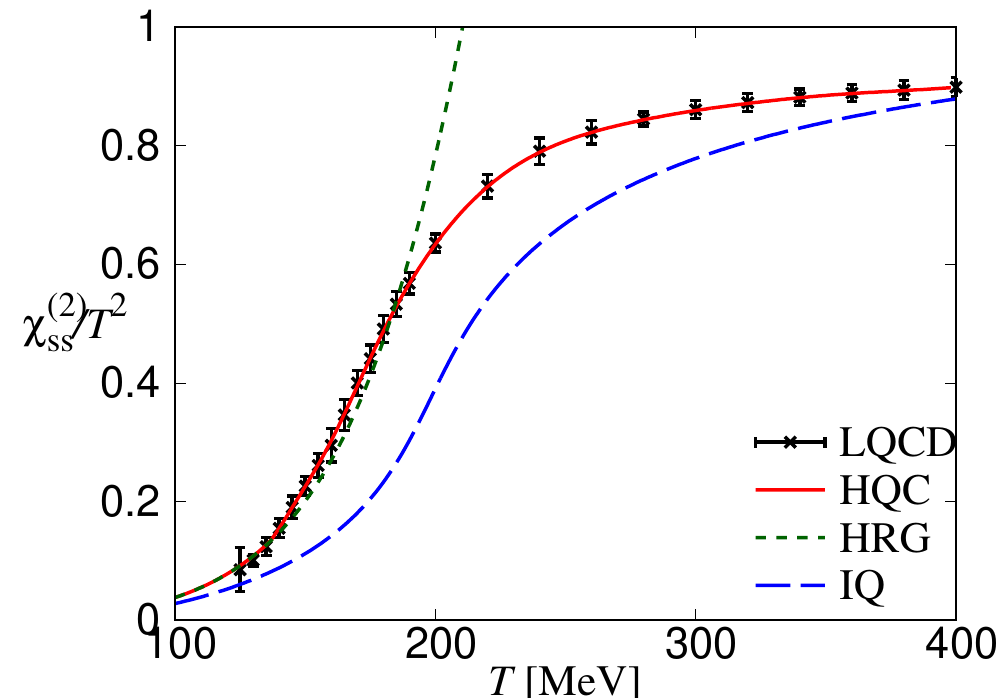}
\end{minipage}\\
\begin{minipage}{0.5\hsize}
\centering
\includegraphics[width=0.79\textwidth]{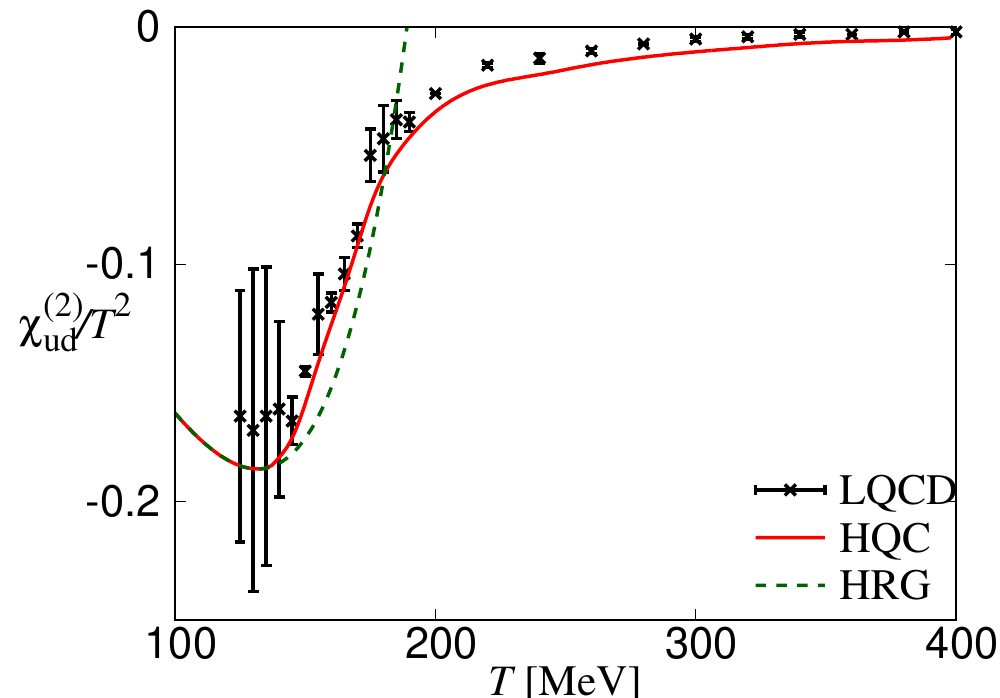}
\end{minipage}
\begin{minipage}{0.5\hsize}
\centering
\includegraphics[width=0.79\textwidth]{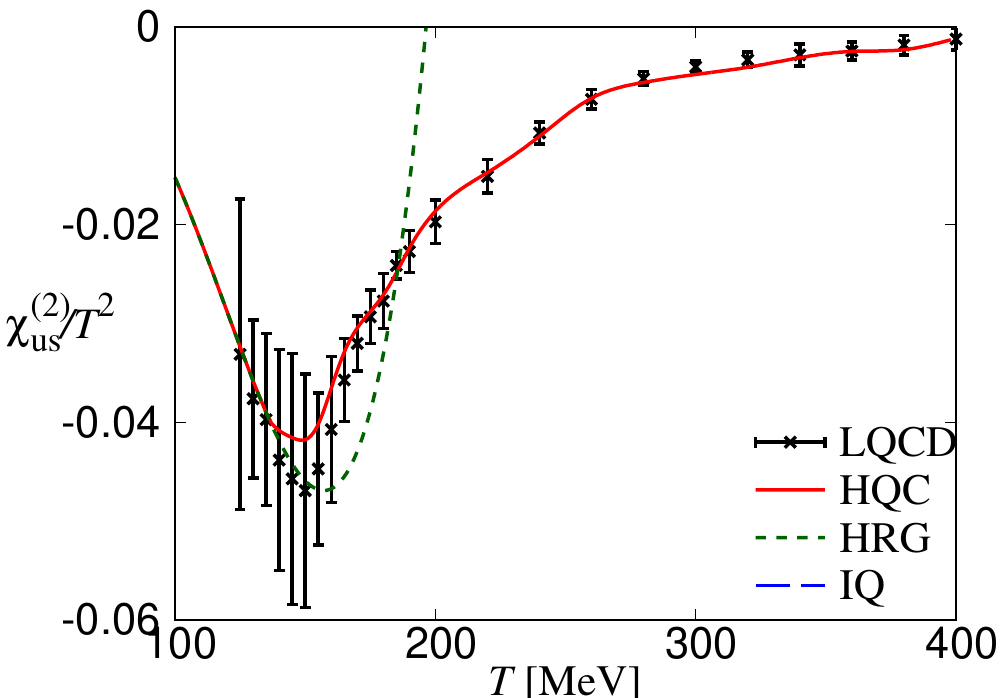}
\end{minipage}
\end{tabular}
\caption{$T$ dependence of diagonal and off-diagonal susceptibilities, 
$\chi_{ff'}^{(2)}$, in the 2+1 flavor system with 
zero chemical potential.  
The HQC result  is drawn by the solid line. 
The dotted line stands for the result of the HRG model, 
the dashed line corresponds to that of the IQ model. 
LQCD data are taken from Ref.~~\cite{Borsanyi_sus_plot}. 
}
\label{ff-susceptibility-2+1}
\end{figure*}

Figure \ref{ff-susceptibility-2+1} shows the flavor 
diagonal and off-diagonal susceptibilities, 
$\chi_{ff'}^{(2)}$, as a function of $T$ in the 2+1 flavor system with 
zero chemical potential. 
The solid line is the result of the HQC model. 
The HQC model should reproduce LQCD data on the $\chi_{ff'}^{(2)}$ 
automatically. This is satisfied. 
Just for comparison, the results of the HRG and IQ models are 
denoted by dashed and dotted lines, respectively. 
As already mentioned in Sec. \ref{II-2+1 flavor system}, 
the IQ model has no contribution for 
the off-diagonal susceptibilities. 
Noting $\chi^{(2)}_{\rm ud} \approx 5 \chi^{(2)}_{\rm us}$, 
one can see from $T$ dependence of the off-diagonal susceptibilities 
that most of hadrons disappear at $T = 400$~MeV. 
The hadron-quark transition thus ends up with $T \approx 400$~MeV.

\subsection{2+1+1 flavor system}
\label{III:2+1+1 system}
\begin{figure}[H]
\centering
\includegraphics[width=0.4\textwidth]{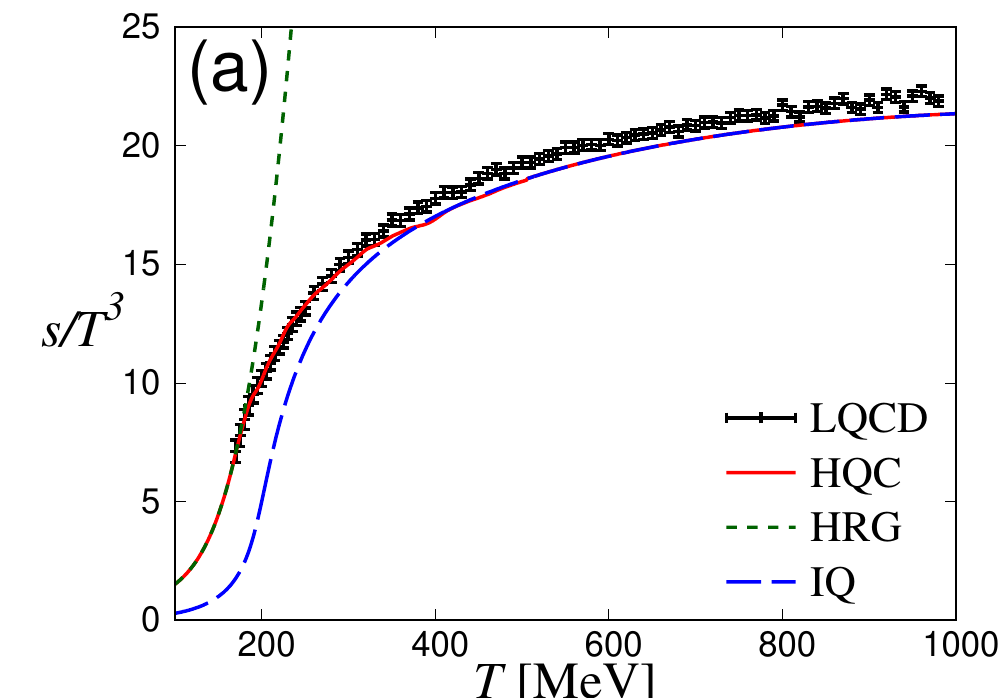}
\includegraphics[width=0.4\textwidth]{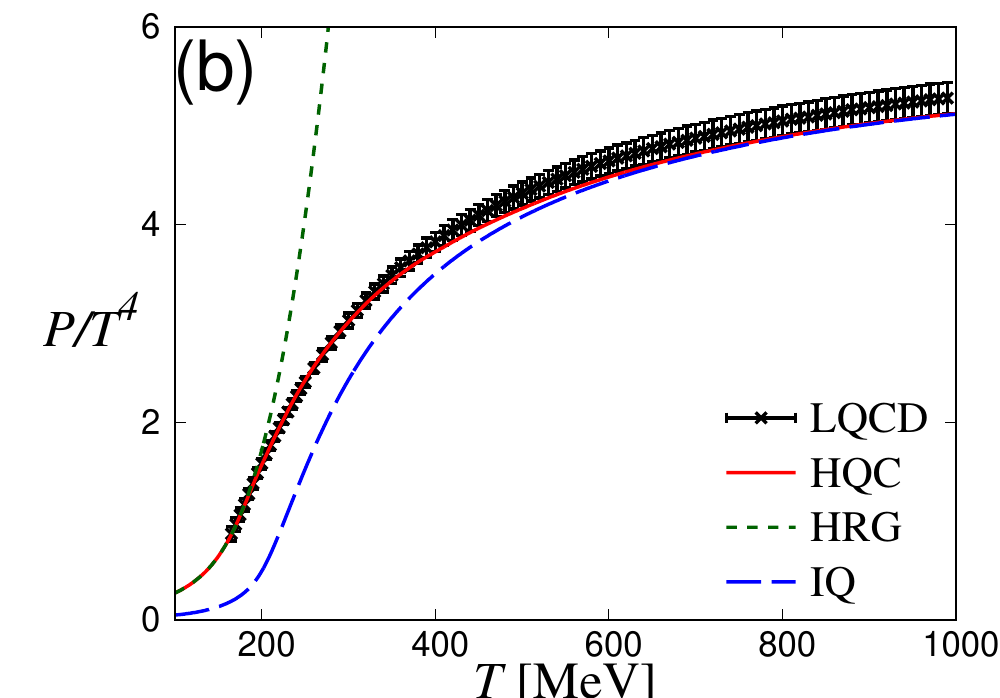}
\caption{$T$ dependence of $s$ and $P$ 
in the 2+1+1 flavor system with zero chemical potential.
The solid line is the result of the HQC model.
LQCD data are taken from Ref.~\cite{Borsanyi:2016ksw}. 
The result of the IQ (HRG) model is denoted by 
a dashed (doted) line. 
}
\label{fig_P 2+1+1}
\end{figure}

Figure \ref{fig_P 2+1+1} shows $s$ and $P$ as a function of $T$ 
in the 2+1+1 flavor system with zero chemical potential. 
Good agreement is seen between the HQC results (solid line) 
and LQCD data. 
Comparing the HQC results with the HRG and IQ ones, 
we can see that the transition region is $170 \lsim T \lsim 400$~MeV. 
The upper limit of the phase transition can be determined 
more clearly with the off-diagonal susceptibilities, as shown 
below.

Figure \ref{ff-susceptibility_2+1+1} shows 
the flavor diagonal and off-diagonal susceptibilities, $\chi_{ff'}^{(2)}$, as a function of $T$ in the 2+1+1 flavor system with zero chemical potential. 
One can see good agreement between LQCD data and the HQC results for 
the 2+1 flavor sector, i.e., $\chi_{\rm uu}^{(2)}$, $\chi_{\rm ud}^{(2)}$, 
$\chi_{\rm ss}^{(2)}$, $\chi_{\rm us}^{(2)}$. 
This supports the statement that 
c quark does not affect the 2+1 flavor subsystem composed of u, d, s quarks, 
together with the fact that $\chi_{\rm uu}^{(2)}$, $\chi_{\rm ud}^{(2)}$, 
$\chi_{\rm ss}^{(2)}$, $\chi_{\rm us}^{(2)}$ in the 2+1+1 flavor system 
are close to the corresponding susceptibilities in the 2+1 flavor system. 
Noting $\chi^{(2)}_{\rm ud} \approx 5 \chi^{(2)}_{\rm us} 
\gg \chi^{(2)}_{\rm uc}$, we can consider from $T$ dependence of 
the off-diagonal susceptibilities
that most of hadrons disappear at $T = 400$~MeV. 
Also for the 2+1+1 flavor system, 
the hadron-quark transition thus ends up with $T \approx 400$~MeV. 
Hence, the transition region 
is $170 \lsim T \lsim 400$~MeV also for the 2+1+1 flavor system 
with zero chemical potential.

The present HQC model neglects $\mu_{\rm c}$-dependence in $f_{\rm H}(T,\{\mu_{\a}\})$, but  reproduces LQCD data qualitatively for $\chi_{\rm cc}^{(2)}$. 
As for $\chi_{\rm uc}^{(2)}$, both LQCD and the HQC model show the correlation 
between u and c quarks is negligible in the transition region 
$170 \lsim T \lsim 400$~MeV.

\begin{figure*}[t]
\begin{tabular}{lr}
\begin{minipage}{0.5\hsize}
\centering
\includegraphics[width=0.8\textwidth]{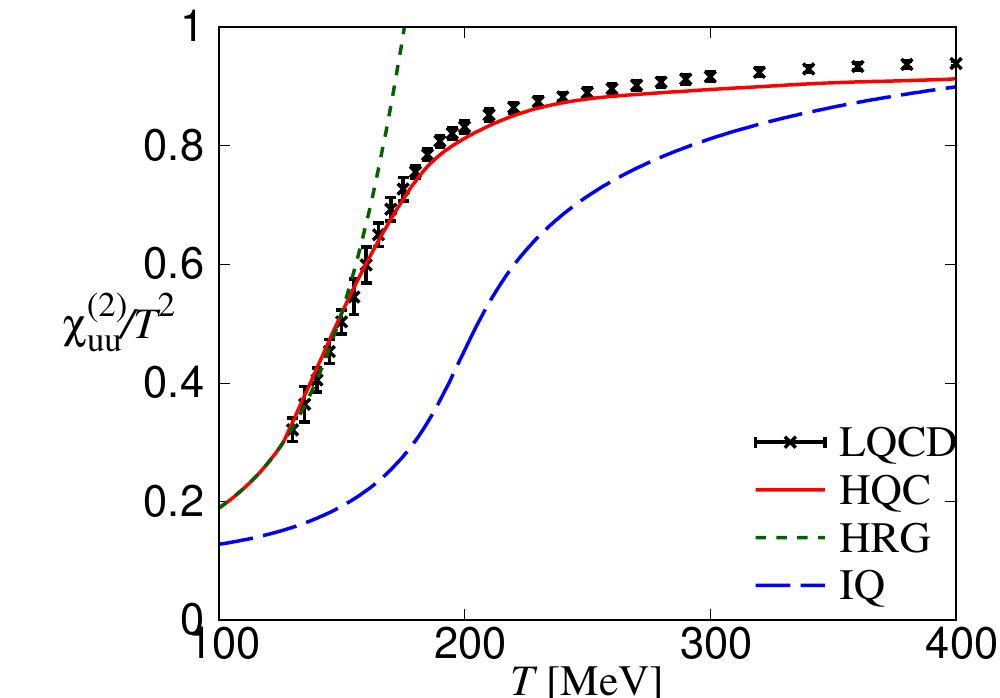}
\end{minipage}
\begin{minipage}{0.5\hsize}
\centering
\includegraphics[width=0.8\textwidth]{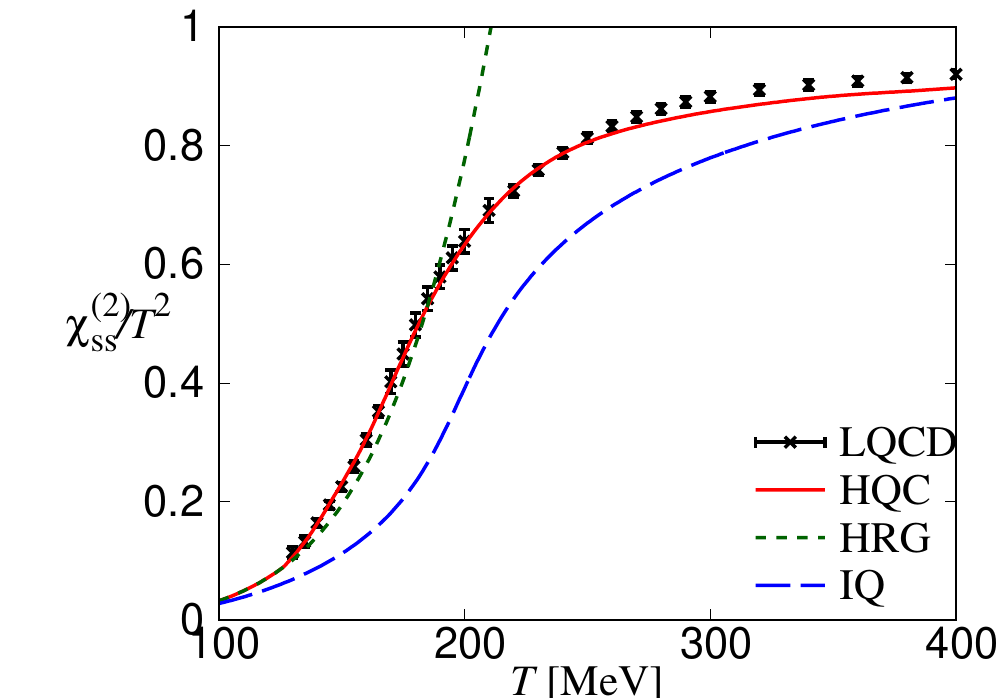}
\end{minipage}\\
\begin{minipage}{0.5\hsize}
\centering
\includegraphics[width=0.8\textwidth]{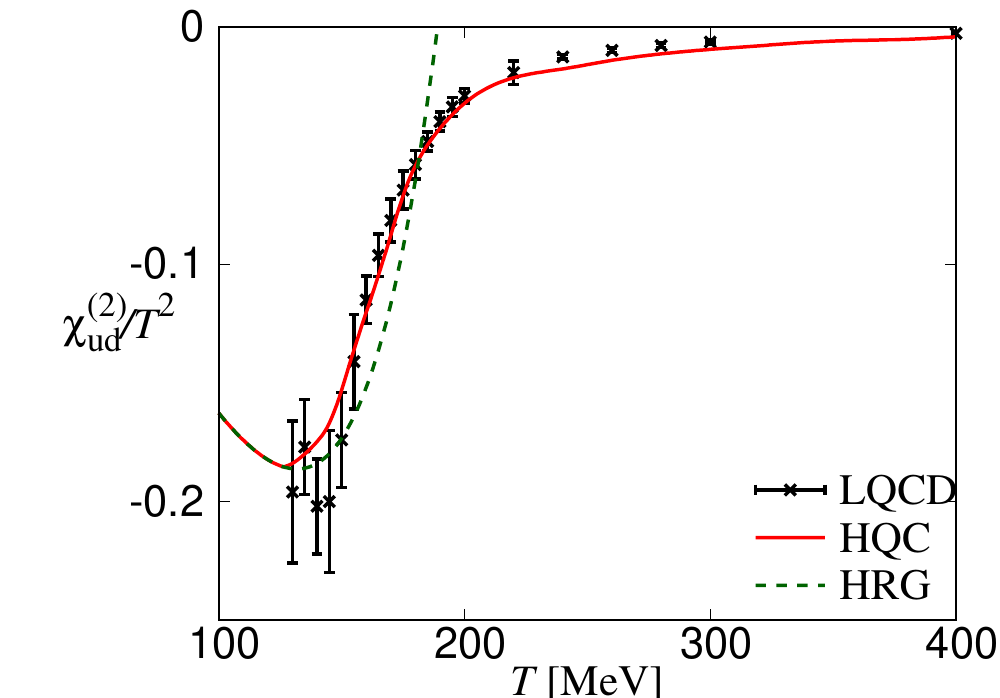}
\end{minipage}
\begin{minipage}{0.5\hsize}
\centering
\includegraphics[width=0.8\textwidth]{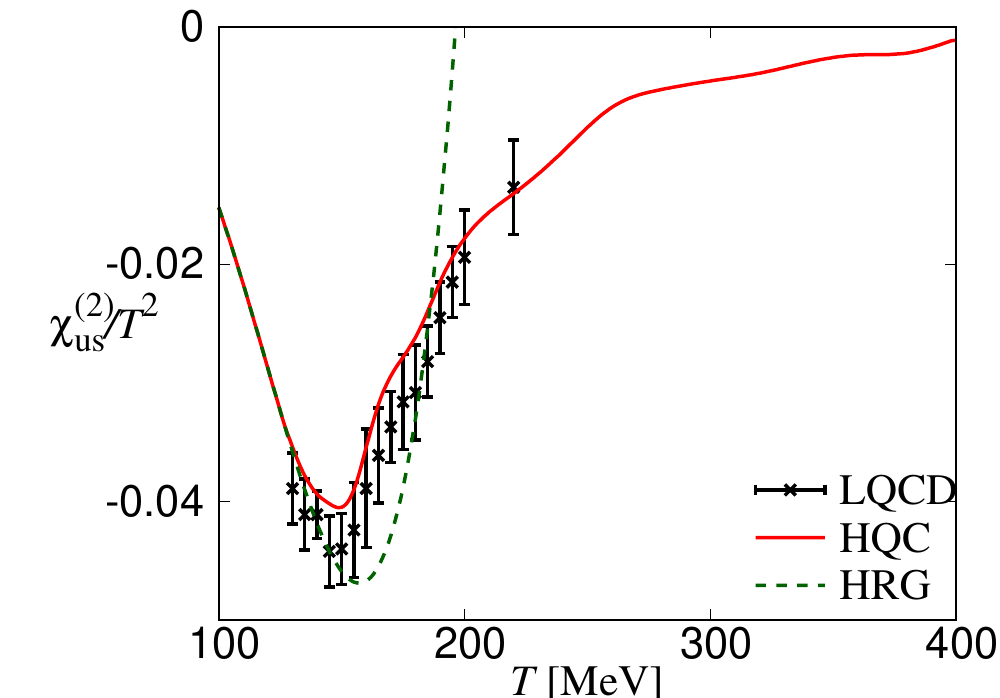}
\end{minipage}\\
\begin{minipage}{0.5\hsize}
\centering
\includegraphics[width=0.8\textwidth]{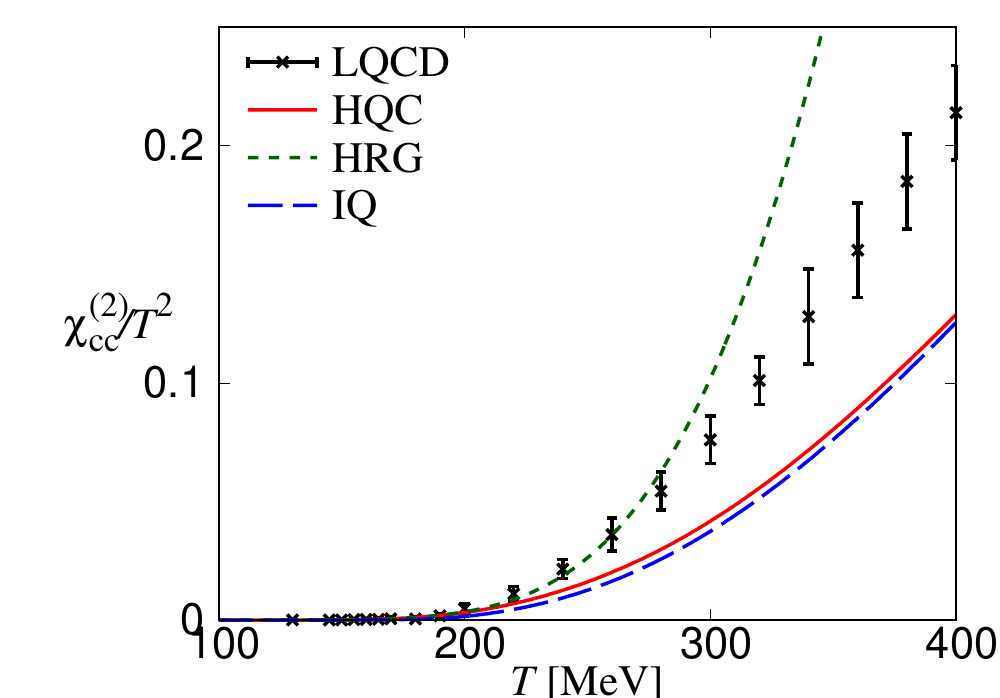}
\end{minipage}
\begin{minipage}{0.5\hsize}
\centering
\includegraphics[width=0.8\textwidth]{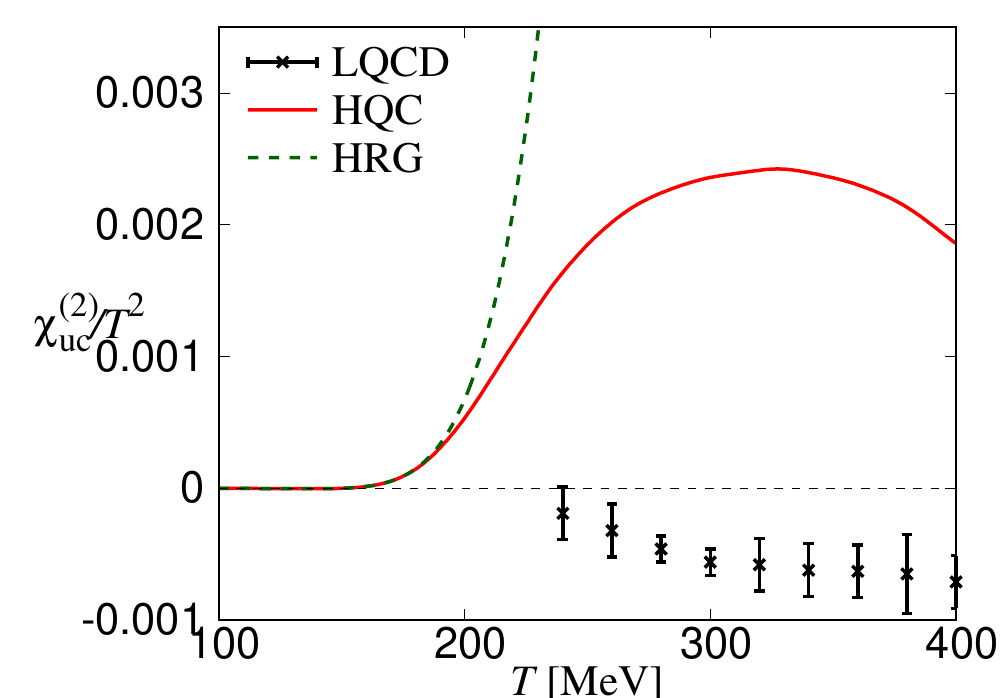}
\end{minipage}
\end{tabular}
\caption{$T$ dependence of flavor diagonal and off-diagonal susceptibilities, 
$\chi_{ff'}^{(2)}$, in the 2+1+1 flavor system with zero chemical potential. 
The solid line denotes the HQC result, while the chain line is the HRG result. 
The dotted line stands for the result of the HRG model, 
the dashed line corresponds to that of the IQ model. 
LQCD (dots) date are taken from Ref~\cite{Borsanyi:2016ksw}. 
}
\label{ff-susceptibility_2+1+1}
\end{figure*}

\begin{figure}[h]
\centering
\vspace{0cm}
\includegraphics[width=0.4\textwidth]{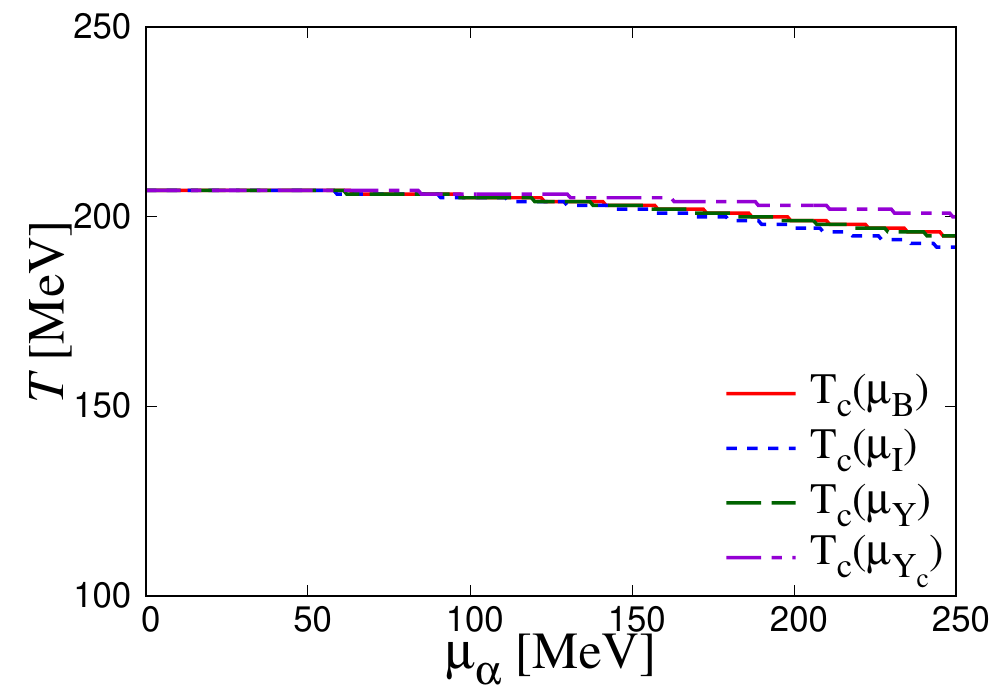}
\caption{Phase diagram in $\mu_{B}$--$T$, $\mu_{I}$--$T$, $\mu_{Y}$--$T$, 
$\mu_{Y_c}$--$T$ planes. }
\label{fig_phase-diagram-BIYYc}
\end{figure}

Finally, we discuss the phase diagram in $\mu_{B}$--$T$, $\mu_{I}$--$T$, $\mu_{Y}$--$T$, $\mu_{Y_c}$--$T$ planes. For this purpose, we have evaluated 
$f^{(2)}_{\rm H,cc}$ from $\chi^{(2)}_{\rm cc}$, but 
$f^{(2)}_{\rm H,cc} \approx f^{(2)}_{\rm H,ss}/100$ around $T=200$~MeV. 
This justifies our assumption that $f^{(2)}_{\rm H,cc}$ is negligible 
in $f_{\rm H}$, if we do not analyze $\chi^{(2)}_{\rm cc}$ itself. 
Here, note that $f^{(2)}_{\rm H,cu}$ is even smaller than $f^{(2)}_{\rm H,cc}$.
Using the assumption 
$f^{(2)}_{\rm H,cu}=f^{(2)}_{\rm H,su}=f^{(2)}_{\rm H,cc}=0$ 
in $f_{\rm H}$, 
we draw the phase diagram in Fig.~\ref{fig_phase-diagram-BIYYc}.
$BIY$ approximate equivalence still persists in the 2+1+1 flavor system, 
but the transition line $T_{c}(\mu_{Y_c})$ in $\mu_{Y_c}$--$T$ plane is 
slightly higher than in the other planes. This is because the difference, $f_{{\rm H},Y_c}^{(2)} - f_{{\rm H},B}^{(2)}$, has the $f^{(2)}_{\rm H,uu}$ that is rather larger than the others around $T=200$~MeV.

\begin{figure}[h]
\centering
\vspace{0cm}
\includegraphics[width=0.4\textwidth]{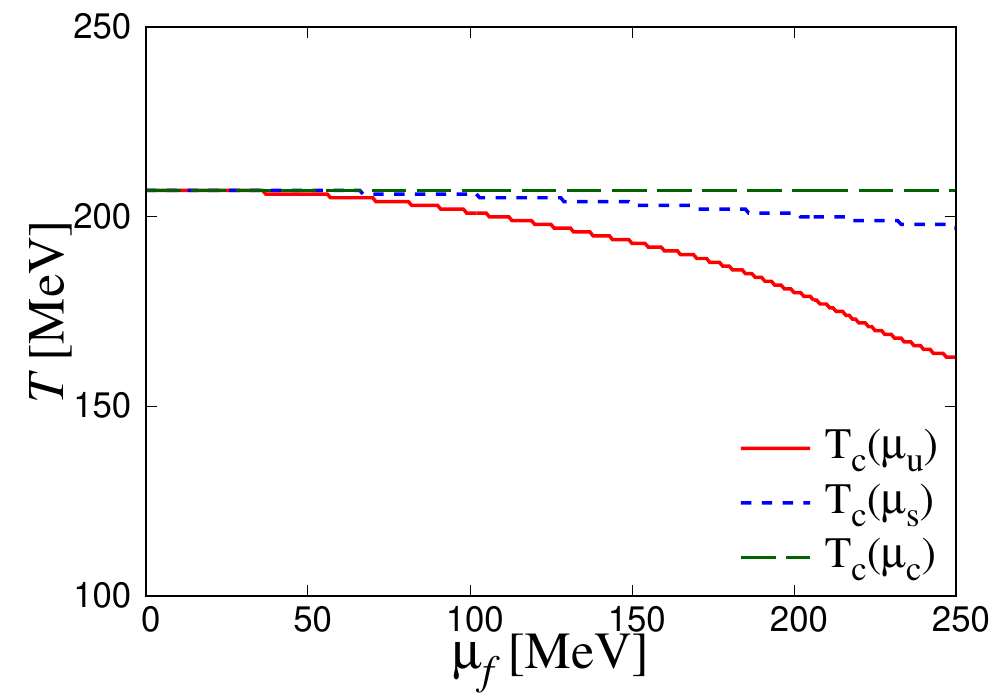}
\caption{Phase diagram in $\mu_{u}$--$T$, 
$\mu_{s}$--$T$, $\mu_{s}$--$T$ planes. 
Note that $T_c(\mu_{\rm u})=T_c(\mu_{\rm d})$ for $\mu_{\rm u}=\mu_{\rm d}$ because of $f_{\rm H,{uu}}^{(2)}(T)=f_{\rm H,{dd}}^{(2)}(T)$ and $f_{\rm H,{us}}^{(2)}(T)=f_{\rm H,{ds}}^{(2)}(T)$. }

\label{phase diagram f 2+1+1}
\end{figure}

Figure \ref{phase diagram f 2+1+1} shows the phase diagram 
in $\mu_{\rm u}$--$T$, $\mu_{\rm d}$--$T$, $\mu_{\rm s}$--$T$, $\mu_{\rm c}$--$T$ planes; 
here note that $T_c(\mu_{\rm u})=T_c(\mu_{\rm d})$ when $\mu_{\rm u}=\mu_{\rm d}$.
In this analysis, the transition line in $\mu_{\rm c}$--$T$ plane
is $T_c(\mu_{\rm c})=207$~MeV, because of $f_{\rm H,{cc}}^{(2)}(T)=0$. 
We can find that $T_c(\mu_{\rm u}) < T_c(\mu_{\rm s}) < T_c(\mu_{\rm c})$ when 
$\mu_{\rm u}=\mu_{\rm s}=\mu_{\rm c}$. 
This result indicates that the hadron-quark transition 
takes place at higher $T$ for heavier quark. 
This is quite reasonable. In the nonrelativistic limit that is 
a good approximation for c quark,  
the QCD partition function is a function of $\mu_{\rm c}-m_{\rm c}$ 
and thereby  
the effect of $\mu_{\rm c}$ is sizably reduced with large $m_{\rm c}$.

\section{Summary}
\label{summary}

We reconstructed the hadron-quark crossover (HQC) model 
of Ref.~\cite{Miyahara:2016din} 
by using new LQCD data~\cite{Borsanyi:2016ksw,Borsanyi_sus_plot} 
in the 2+1 flavor system, and we drew the phase diagram 
in the 2+1 and 2+1+1 flavor systems 
through analyses of the EoS and the susceptibilities.

The HQC model is defined by Eq.~\eqref{s_Hybrid_def} 
in which $s_{\rm H}(T,\{ \mu_\a \})$ 
is calculated by the hadron resonance gas (HRG) model valid in low $T$ 
and $s_{\rm Q}(T,\{ \mu_\a \})$ 
is by the independent quark (IQ) model reasonable in 
high $T$, where $\{ \mu_\a \}=(\mu_B, \mu_I, \mu_Y)$. 
As mentioned above, the present version of the IQ model 
is rather reliable, since it explains 
LQCD data on the EoS in $400 \lsim T \leq  500$~MeV 
for the 2+1system and in $400 \lsim T \leq  1000$~MeV the 2+1+1system 
where LQCD data is consistent with NNLO HTL 
perturbation~\cite{Haque:2013sja}.

The hadron-production probability $f_{\rm H}(T,\{ \mu_\a \})$ 
was determined from 
LQCD data on $s$ and the susceptibilities 
$\chi_{\gamma}^{(2)}$ for $\gamma=B, I, Y,BY$ in the 2+1 flavor system. 
Hence, the present HQC model automatically reproduces LQCD data on the EoS and 
the $\chi_{ff'}^{(2)}$ in the 2+1 flavor system. 
In particular, the off-diagonal susceptibilities $\chi_{ff'}^{(2)}$ 
($f \neq f'$) are good indicators to see 
how hadrons survive as $T$ increases, since the IQ  model hardly
contributes to the off-diagonal susceptibilities. 
In fact, the off-diagonal susceptibilities show that most of hadrons disappear 
above $T=400$~MeV. 
In practice, the upper limit of the transition region is clearly determined 
by the off-diagonal susceptibilities. 
We then determined, from $T$ dependence of $f_{\rm H}^{(0)}(T)$ and 
the off-diagonal susceptibilities,
that the transition region is $170 \lsim T \lsim 400$~MeV
for the 2+1 flavor system with zero chemical potential.

In the present paper, we defined the hadron-quark  transition 
temperature by $f_{\rm H}(T,{\{\mu_\a \}})=1/2$. 
For the 2+1 flavor system with zero chemical potential, 
the transition temperature is $T_c^{(f_{\rm H})}=207$~MeV 
and somewhat larger than LQCD result $T_c^{(\Phi),{\rm LQCD}}=170 \pm 7$~MeV. This result $T_c^{(f_{\rm H})}=207$~MeV is common between 
the 2+1 and 2+1+1 flavor systems, because  
$f_{\rm H}^{(0)}(T)$ is the same between the two systems.

As mentioned above, the HQC model well simulates LQCD data 
on the EoS and the $\chi_{ff'}^{(2)}$ 
in the 2+1 flavor system. We then drew the phase diagram 
in $\mu_{B}$--$T$, $\mu_{I}$--$T$, $\mu_{Y}$--$T$ planes. 
We found ``$BIY$ approximate equivalence": Namely,  
the transition lines $T_c(\mu_{\a})$ are almost identical in these planes, 
when the $\mu_{\a}$ are less than 250~MeV.
The relation $T_c(\mu_{B}) \approx T_c(\mu_{Y})$ comes from the fact that 
the s-quark contribution is small in the $f_{{\rm H},\gamma}^{(2)}(T)$ for $\gamma=B, I, Y, BY$. 
The relation $T_c(\mu_{B}) \approx T_c(\mu_{I})$ may be  
a remnant of the fact 
that in the 2 flavor system $T_c(\mu_{B})=T_c(\mu_{I})$ when 
these are expressed 
up to $(\mu_{B}/T)^2$ and $(\mu_{I}/T)^2$~\cite{Ejiri:2007ga}.

The HQC model was applied to 
the 2+1+1 flavor system without changing the $f_{\rm H}$. 
The HQC model well reproduces LQCD data on the EoS and 
the $\chi_{ff'}^{(2)}$ for $f,f=$u, d, s. 
This result indicates that transition region at ${\{\mu_\a \}}=0$ 
is $170 \lsim T \lsim 400$~MeV also for the 2+1+1 flavor system, 
since between the 2+1 and 2+1+1 flavor systems 
the $f_{\rm H}^{(0)}$ is the same and the $\chi_{ff'}^{(2)}$ for $f,f=$u, d, s 
are close to each other. 
In addition, $T$ dependence of $\Phi$ is almost identical 
between the 2+1 and 2+1+1 flavor systems. These results show that 
c quark does not affect the 2+1 flavor subsystem composed of u, d, s quarks. 
This statement is supported by the fact that 
$\chi_{\rm ud}^{(2)} \approx 5 \chi_{\rm us}^{(2)} \gg \chi_{\rm uc}^{(2)}$ 
in the transition region.

The present HQC model has no $\mu_{\rm c}$-dependence in $f_{\rm H}(T,\{\mu_{\a}\})$, but  reproduces LQCD data qualitatively for $\chi_{\rm cc}^{(2)}$. 
As for $\chi_{\rm uc}^{(2)}$, both LQCD and the HQC model show the correlation 
between u and c quarks is negligible in the transition region 
$170 \lsim T \lsim 400$~MeV.

Finally, we plotted the phase diagram both 
in $\mu_{B}$--$T$, $\mu_{I}$--$T$, $\mu_{Y}$--$T$, $\mu_{Y_c}$--$T$ planes and 
in $\mu_{\rm u}$--$T$, $\mu_{\rm d}$--$T$, $\mu_{\rm s}$--$T$, $\mu_{\rm c}$--$T$ planes.
$BIY$ approximate equivalence still persists in the 2+1+1 flavor system, 
but the transition line $T_{c}(\mu_{Y_c})$ in $\mu_{Y_c}$--$T$ plane is 
slightly higher than in the other planes.
We also found that $T_c(\mu_{\rm u}) < T_c(\mu_{\rm s}) < T_c(\mu_{\rm c})$ when 
$\mu_{\rm u}=\mu_{\rm s}=\mu_{\rm c}$. 
This result indicates that the hadron-quark transition 
takes place at higher $T$ for heavier quark. 
This is quite natural. In the nonrelativistic limit that is 
a good approximation for c quark,  
the QCD partition function is a function of $\mu_{\rm c}-m_{\rm c}$, so that 
the effect of $\mu_{\rm c}$ is greatly reduced with large $m_{\rm c}$.

\noindent
\begin{acknowledgments}
The authors thank Atsushi Nakamura, Kouji Kashiwa, Junichi Takahashi, Junpei Sugano, Shuichi Togawa ,Yuhei Torigoe and Takehiro  Hirakida for useful discussions. 
M. Y. and H. K. were supported
by Grant-in-Aid for Scientific Research (No. 26400278 and No. 26400279) 
from the Japan Society for the Promotion of Science (JSPS). 
\end{acknowledgments}

\appendix


\end{document}